\documentclass[lettersize,journal,10pt,times]{IEEEtran}
\usepackage{cite}
\usepackage{amsmath,amssymb,amsfonts}
\usepackage{algorithmic}
\usepackage{textcomp}
\usepackage{stfloats}
\usepackage{url}
\usepackage{verbatim}
\usepackage{graphicx}
\usepackage{tabularx}
\usepackage{multirow}
\usepackage{subcaption}
\usepackage{xcolor}
\usepackage{amsthm,mathtools,nccmath}
\usepackage{xfrac}
\usepackage{dutchcal}
\usepackage{ctable} 
\theoremstyle{definition}

\usepackage{MnSymbol}
\usepackage{enumitem} 
\usepackage{hyperref}
\hypersetup{
  colorlinks=true,
  linkcolor=blue,
  urlcolor=cyan,
  citecolor=black,
}

\def\BibTeX{{\rm B\kern-.05em{\sc i\kern-.025em b}\kern-.08em
    T\kern-.1667em\lower.7ex\hbox{E}\kern-.125emX}}
\usepackage{balance}
\usepackage[dont-mess-around]{fnpct}   
\usepackage{bigfoot}
\begin{document}
\setlength{\skip\footins}{6pt}
\title{BERP: A Blind Estimator of Room Parameters for Single-Channel Noisy Speech Signals}
\author{%
\IEEEauthorblockN{%
Lijun Wang\,\IEEEauthorrefmark{1}~\IEEEmembership{Student Member, IEEE},
Yixian Lu\,\IEEEauthorrefmark{1}, 
Ziyan Gao\,~\IEEEmembership{Member, IEEE}, Kai Li\,,
Jianqiang Huang\,,
Yuntao Kong\,,
and Shogo Okada\,\IEEEauthorrefmark{2}~\IEEEmembership{Member, IEEE}
}
\thanks{L. Wang, Z. Gao, K. Li, J. Huang, Y. Kong, and S. Okada are with School of Information Science, Japan Advanced Institute of Science and Technology, Ishikawa, Japan. email:\,\{lijun.wang, ziyan-g, kai-li, jq.huang, okada-s, yuntao.kong\}@jaist.ac.jp. Y. Lu is with ACES, Inc., Tokyo, Japan. email:\, mcluoc@gmail.com. This paper has supplementary downloadable material available at http://ieeexplore.ieee.org., provided by the author. The material includes details of the hybrid approach. Contact okada-s@jaist.ac.jp for further questions about this work. This work was partially supported by the Japan Society for the Promotion of Science (JSPS) KAKENHI (grant number 23H03506) and JST CRONOS (grant number JPMJCS24K7).}
\thanks{\IEEEauthorrefmark{1}Equal contribution.}
\thanks{\IEEEauthorrefmark{2}Corresponding author.}
\thanks{\footnotemark{1}\url{https://github.com/Alizeded/BERP}.}
}

\markboth{Journal of \LaTeX\ Class Files,~Vol.~xx, No.~xx, xx.~xxxx}%
{How to Use the IEEEtran \LaTeX \ Templates}

\maketitle
\begingroup\renewcommand\thefootnote{\textsection}
\endgroup
\begin{abstract}
Room acoustical parameters (RAPs), room geometrical parameters (RGPs) and instantaneous occupancy levels are essential metrics for parameterizing the room acoustical characteristics (RACs) of a sound field around a listener's local environment, offering comprehensive indications for various applications. 
Current blind estimation methods either fail to cover a broad range of real-world acoustic environments in the context of real background noise or estimate only a few RAPs and RGPs from noisy single-channel speech signals. In addition, they are limited in their ability to estimate the instantaneous occupancy level. In this paper, we propose BERP, a new universal approach to blindly estimate RAPs, RGPs, and occupancy levels. It consists of two modules: one for RAPs and RGPs and another for occupancy levels. For the former task, we use a shared room feature encoder that combines attention mechanisms with convolutional layers to learn common features across room parameters, and multiple separate parametric predictors for continuous estimation of each parameter in parallel. The combination of attention and convolutions enables the model to capture acoustic features both locally and globally from speech, yielding more robust and multitask generalizable common features. Separate predictors allow the model to independently optimize for each room parameter to reduce task learning conflict and improve per-task performance. This architecture enables universal and efficient estimation of room parameters while maintaining satisfactory performance. For occupancy level estimation, we reuse the identical encoder with a classification head, exploiting the encoder’s strong sequence‑feature extraction. To evaluate the effectiveness of the proposed approach, we compile a task-specific dataset from several publicly available datasets, including synthetic and real reverberant recordings. The results reveal that BERP achieves state-of-the-art (SOTA) performance and excellent adaptability to real-world scenarios. The code, datasets, and weights are available on GitHub \footnotemark.
\end{abstract}
\begin{IEEEkeywords}
    Room acoustics, Blind estimation, Room acoustic parameters, Deep learning, Multitask learning, Attention
\end{IEEEkeywords}

\section{Introduction}
\label{sec:intro}
\IEEEPARstart{R}{oom} acoustical characteristics (RACs) characterize the acoustical properties of a room through which people perceive sounds in an enclosure.
RACs determine how intelligibly and clearly people perceive sounds in an auditory space enclosed by walls, ceilings, and furniture. Local RACs, which refer to the RACs perceived within the listener's local surroundings, are widely employed in speech enhancement, hearing aids, immersive audio, context-aware renderings (such as mixed and augmented reality), public address systems, and robotics. Compared with global RACs, which reflect the stationary attributes of room acoustics, the dynamic parameterization of local RACs poses a significant challenge in room acoustics, given the interference caused by background environmental noise and its position-dependent attributes.

Subjective listening tests can be conducted to assess local RACs in a listener's surroundings. However, conducting a subjective test is expensive and time-consuming, and more importantly, subjective results are significantly influenced by individual differences such as age, hearing ability, and cultural background. Instead, room acoustical parameters (RAPs), such as the reverberation time and the speech intelligibility index, are usually used as the objective indices to assess the local RACs. Room geometrical parameters (RGPs), such as the room volume, sound source distance and direction of arrival (DOA), are also important for evaluating local RACs, as they have critical applications in spatial audio rendering, intelligibility assessments in a room, separation of sound sources, audio navigation systems and speech enhancement \cite{jenrungrot_cone, Chazan2019MultiMicrophoneSS, jot_augmented_2016, vanderwerff2003what, doa_app_roomanalysis, ssl_app_navi, doa_app_speechenhance}. Consequently, RAPs and RGPs can be used together to model local RACs to offer clear and comprehensive indications for various applications, such as room acoustical assessment \cite{IEC60268, Alcons, ISO3382, doa_app_roomanalysis, kinoshita2013}, speech enhancement \cite{Lior2024, Morgenstern_2017, doa_app_speechrecog}, hearing aids \cite{Tom2024, REYNDERS2024, Fogerty2020, EURICH2019122, vanderwerff2003what, doa_app_speechrecog}, sound source separation \cite{jenrungrot_cone, Chazan2019MultiMicrophoneSS}, spatial audio rendering \cite{yenduri2024spatial}, context-aware rendering in extended reality (XR) and augmented reality (AR) \cite{Hugues2024, jot_augmented_2016, Joonas2014}, public address systems \cite{Armin2020, Dong2018}, and robotics \cite{ssl_app_interact}. Another important metric is the instantaneous occupancy level, which quantifies how many people are present near the listener at a given time; this is an essential consideration in demand-driven HVAC (heating, ventilation, and air conditioning) control systems for smart buildings \cite{occu_hvac, occu_hvac_2}.

Several RAPs have been investigated and standardized \cite{kuttruff2016room, Alcons, IEC60268, ISO3382}, including the speech transmission index (STI), the percentage articulation loss of consonants ($\%\rm{AL}_{\rm{cons}}$), reverberation time ($T_{60}$), early decay time (EDT), clarity ($C_{80}$ / $C_{50}$), definition ($D_{50}$) and center time ($T_s$). RAPs can be directly derived from the measured RIR. However, measuring RIR requires the exclusion of people in an enclosure, which is impractical for public spaces, as it involves the use of high-energy sounds\cite{blindEst_wang, Suradej}. Furthermore, it is limited in capturing the dynamics of local RACs, which vary depending on the locations, arrangements, and quantities of objects and occupants present. The RAPs measured using specific standards may differ from noncompliant measures employed within the same enclosure. Moreover, RGPs may be derived from the measured RIR, but they encounter the same aforementioned issues. Therefore, some blind estimation methods have been proposed to estimate RAPs and RGPs, particularly in public spaces where people cannot be excluded.

In general, blind estimation is a challenging task because it is an ill-posed inverse problem without prior knowledge about systems and inputs. In this case, conventional blind estimation methods use mathematical derivations to create an explicit map between observed signals and systems\cite{Kendrick, Keshavarz, blindEst_wang}, whereas current blind estimation methods typically employ deep neural networks (DNNs) to map observed signals to room parameters owing to their superior performance over conventional analytical methods\cite{ACE2016}, such as convolutional neural networks (CNNs) \cite{meng2025, FullCNN_2023, ZHENG2022110901, Srivastava2021BlindRP, FullCNN_2019, FullCNN_2018}, convolutional recurrent neural networks (CRNNs) \cite{meng2025,CRNN_2021, CRNN_2020, Neri_2024, dist_est_kushawaha}, and hybrid approaches\cite{Suradej}.

\begin{figure}
\vspace{-1.0em}
    \centering
    \includegraphics[width=0.63\linewidth]{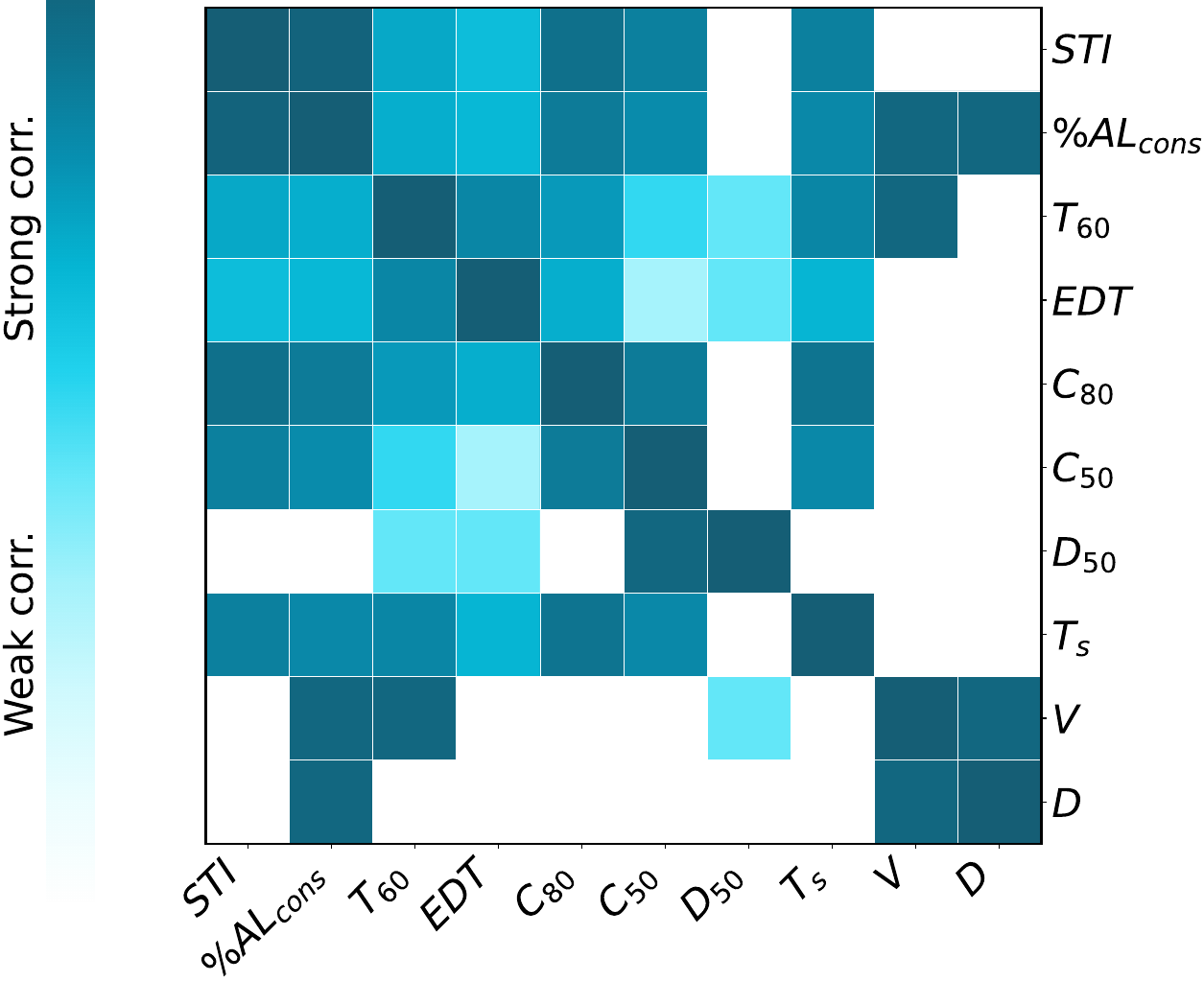}
    \caption{Illustration of mutual relatedness of RAPs and RGPs.}
    \label{fig:motiv}
    \vspace{-1.8em}
\end{figure}

Despite these advances, two key limitations remain. First, most methods estimate only a small subset of parameters (e.g., one, two, or four parameters), which is insufficient for fully assessing local RACs. None of them can simultaneously predict the full set of RAPs and RGPs. Although hybrid approaches can theoretically handle multiple parameters at once, their accuracy is limited by how well the RIR model matches real-world conditions\cite{Suradej}, and they fail to address RGPs. In this direction, we propose a solution to address this issue, i.e., blind estimation of a full set of RAPs and RGPs. We formulate this task as a multitask learning problem, treating each room parameter as a separate task. RAPs and RGPs have mutual relatedness, as they share common reverberation and geometric information within the RIR. Cerdá analyzed the correlations among RAPs\cite{CERDA200997} and studies on Sabine's formula and the critical distance of intelligibility ($D_{ci}$) revealed a correlation between room volume and reverberation time\cite{kuttruff2016room, vanderwerff2003what}. \cite{vanderwerff2003what} also explored the relationships among $\%\text{AL}_{\text{cons}}$, room volume, and sound source distance. Figure\,\ref{fig:motiv} illustrates the mutual relatedness among the room parameters as shown in \cite{CERDA200997, kuttruff2016room}, and \cite{vanderwerff2003what}. On the one hand, exploiting mutual relatedness among room parameters may improve per-task performance, particularly for harder tasks or tasks with limited training examples, since task relatedness promotes knowledge transfer to allow more effective learning \cite{Shai2003, ruder2017overviewmultitasklearningdeep}. On the other hand, adding more room parameters as related tasks provides auxiliary task learning benefits to further improve the performance of one or more primary tasks, especially when those tasks have no strong relatedness\cite{ruder2017overviewmultitasklearningdeep, Wang2022}. Moreover, it often suffers from task learning conflicts\cite{Wang2022}.
One possible solution is to focus on model design to reduce task training conflicts and improve multitask performance.

Second, a key challenge lies in the blind estimation of the instantaneous occupancy level. While RAP and RGP estimations have received considerable attention, estimation of the instantaneous occupancy level has been largely overlooked; this may be because RAP and RGP estimations typically assume a single-source speech signal with background noise, whereas the occupancy level estimation assumes a complex, overlapping speech signal, which seems to be significantly more challenging. To simplify the task, some existing methods treat occupancy as a stationary status to estimate a single value\cite{occu_hvac,Chen2017AnIO}. However, occupancy in the real world changes dynamically over time, making it more appropriate to predict a time sequence. Moreover, these methods may not provide a meaningful reflection of performance, as the F1 score considers any prediction that is not exactly correct as entirely wrong, which may not capture how close a predicted count is to the true count. Instead, directly measuring how closely the predicted count aligns with the ground truth by using distance metrics (e.g., mean absolute error (MAE)) is much more precise. To our knowledge, predicting instantaneous occupancy levels over time, particularly considering room acoustic settings, has received little attention. If possible, we expect this task to be unified with RAP and RGP estimations via a universal methodology. Owing to the scarcity of publicly available data for instantaneous occupancy-level estimation, we introduce a data preparation approach specifically designed to address this gap.

\begin{figure*}
\vspace{-1.0em}
    \centering
    \includegraphics[width=0.95\textwidth]{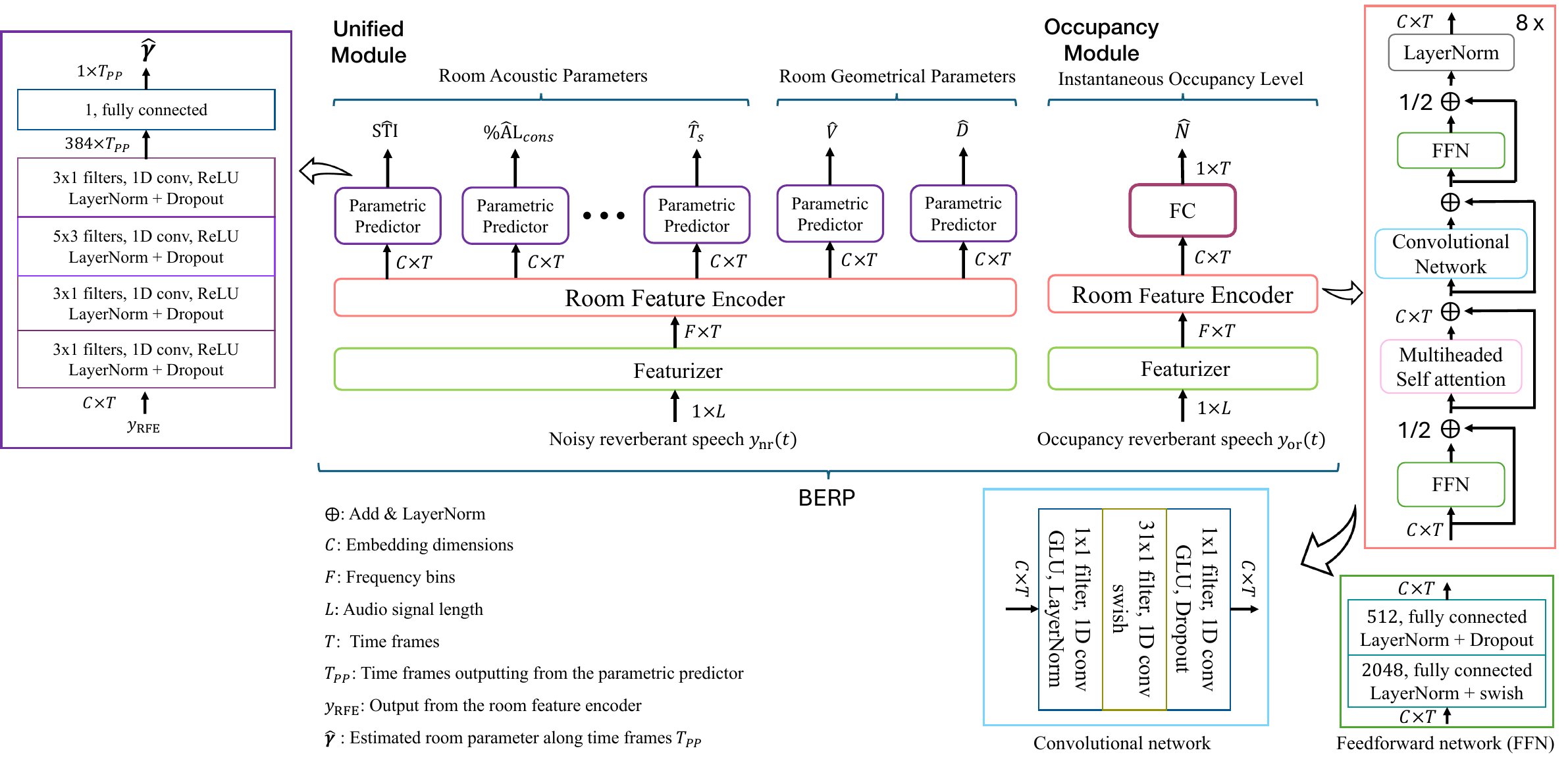}
    \vspace{-0.0em}
    \caption{Overview of the BERP. The input includes the observed noisy and multitalker reverberant speech signals ($y_{nr}(t)\in \mathbb{R}^{1\times L}$ and $y_{or}(t)\in \mathbb{R}^{1\times L}$). At inference time, we sum up and average $\hat{\boldsymbol{\gamma}} \in \mathbb{R}^{1 \times T_{PP}}$ to obtain a single value $\hat{\gamma}\in \mathbb{R}^{1}$. Section\,\ref{ssec:framework_arch} details the featurizer and architectures of the room feature encoder and the parametric predictor.}
    \label{fig:block_diagram}
    \vspace{-1.5em}
\end{figure*}

Motivated by these findings, we propose a new approach (BERP) that tackles two distinct tasks: (1) joint estimation of RAPs and RGPs (namely unified module) and (2) estimation of occupancy levels (namely occupancy module), as shown in Fig.\,\ref{fig:block_diagram}. More specifically, for the joint estimation of RAPs and RGPs, we design a modular architecture that consists of a shared encoder and multiple task-specific predictors. Capturing acoustic features locally and globally together from noisy reverberant speech with complicated waveforms may enable the encoder to learn more robust and multitask-generalizable common features since acoustical information spreads the observed speech signal\cite{blindEst_wang}. Therefore, the shared encoder combines attention and CNNs to grasp features in a time-frequency representation both globally and locally\cite{Wu2020LiteTW, conformer}. The task-specific predictor, which acts as a regressor, predicts a continuous estimation of a room parameter from learned common features. Regarding the blind estimation of the occupancy level, inspired by automatic speech recognition that classifies tokens in each time frame, we formulate this task as a sequence classification problem. Because this problem involves time series data and given that the aforementioned encoder excels in extracting features for a sequence, we directly apply the same encoder architecture. Reusing the encoder, we estimate RAPs, RGPs, and the occupancy level within the same methodology. 

\vspace{-1.0em} 
Currently, except for the occupancy level, BERP assumes a single-source speech signal with background noise as the observed signal. This assumption, while effective for many scenarios, may not fully capture the complexity of real-world acoustic environments. More discussion of current limitations and future work is included in Section\,\ref{sec:conclu}.

Our work makes three important contributions to the frontier of temporary knowledge.
\begin{itemize}
    \item A new universal blind estimation approach is proposed to blindly estimate room parameters and occupancy levels, which achieves state-of-the-art (SOTA) performance.
    \item Signal models for estimating observed speech signals, especially for the instantaneous occupancy level, and corresponding data synthesis pipelines are proposed.
    \item Extensive experiments with synthetic and real data are performed to validate the proposed approach, including in-depth ablation studies regarding model architecture and training data. 
\end{itemize}

The remainder of the manuscript is organized as follows.  Section\,\ref{sec:related_work} provides a literature reviews of state-of-the-art works. Section\,\ref{sec:rp} briefly reviews the room parameters, including the RAPs, RGPs, and instantaneous occupancy level. Section\,\ref{sec:gen_train_data} specifies the data generation pipeline for the proposed method. The proposed approach is introduced in Section\,\ref{sec:proposed}. The experiments are described in Section\,\ref{sec:expres}. We discuss the findings and conclude in Section\,\ref{sec:conclu}.

\section{Related works}
\label{sec:related_work}
\textbf{Blind estimation for RAPs and RGPs.}
DNNs currently lead the field of blind room parameter estimation, typically transforming time-frequency representations of noisy signals into target parameters via CNN-based or CRNN-based architectures. A fully convolutional approach (based on Gamper and Tashev’s work \cite{FullCNN_2018}) has been used to jointly estimate $T_{60}$ and room volume \cite{FullCNN_2023, FullCNN_2019}. López et al. \cite{CRNN_2021} and Callens et al. \cite{CRNN_2020} introduced the CRNN to estimate STI, $T_{60}$, $C_{80}$ and $C_{50}$, outperforming the best model in the ACE challenge \cite{ACE2016}. The CRNN has also been applied to source distance estimation via single- or two-channel speech \cite{Neri_2024, dist_est_kushawaha}. 

\vspace{-0.07em}
Further advances include combining CNNs with MLPs to jointly estimate $T_{60}$, volume, and room surface using single- or two-channel noisy speech \cite{Srivastava2021BlindRP} and employing gating mechanisms within CNNs for $T_{60}$ estimation \cite{ZHENG2022110901}. Duangpummet et al.\cite{Suradej} proposed a hybrid model that integrates RIR modeling with CNNs, allowing concurrent estimation of STI, $T_{60}$, $C_{80}$, $D_{50}$, and $T_s$. Attention has also gained traction: Wang et al.\cite{wang2024, Wang2024volume} explored pure attention mechanisms for estimating $T_{60}$ and volume. Meng et al.\cite{meng2025} estimated $T_{60}$ and $C8_{50}$ for ambisonics recordings using the CNN and the CRNN architectures. Moreover, new training strategies are emerging beyond supervised end-to-end approaches. Philipp et al.\cite{Philipp2024} introduced a task-agnostic embedding method to replace direct supervised learning.

\vspace{-0.15em}
Most current methods can estimate only a small subset of room parameters, which is insufficient for comprehensively describing local RACs. We argue that this limitation may result from their architectural design. They use a single architecture in which all tasks share the same neural network parameters and produce multiple room parameters from one output layer. This setup may lead to serious task-training conflicts and undermine per-task performance, since each task is forced to compete within the same optimization space. One task’s gradient update may severely disrupt another, causing gradient conflicts and making it difficult for the model to find an effective optimization path that satisfies all tasks from the start\cite{Yu2020}. As a result, when estimating many parameters (such as ten), more serious task conflicts may significantly undermine their potential to achieve satisfactory per-task performance.

\vspace{-0.22em}
Inspired by these explorations, we develop a new model. In essence, our method differs algorithmically in how it manages shared and task-specific optimizations to improve multitask performance and reduce task conflicts\cite{Sener2018, Wang2022}. For this purpose, we decompose the model into two parts: a shared encoder that captures common features and multiple task-specific predictors for given room parameters. This architecture reduces interference by preventing multiple tasks from vying for the same output layer, which enables each predictor to specialize in its task, while the encoder focuses on learning common features that are more beneficial across all tasks. More specifically, we combine attention and convolutions to design the shared encoder. The task-specific predictor consists of convolutions that are well suited for the regression task. There are three key motivations behind this design. First, attention enables our encoder to capture dependencies across any two positions, incorporating both long-range and global contextual information; this is crucial in multitask settings, where different tasks may require attention to various segments of the input. This encoder can extract high-level features to feed into multiple predictors for different tasks, reducing parameter redundancy and fostering knowledge sharing between tasks\cite{liu-etal-2019-multi}. In contrast, traditional CNNs or RNNs used in baselines struggle to capture such long-range dependencies and are less adaptable to flexible multitask outputs. Next, using a shared encoder provides an inductive bias that learns valuable features across multiple tasks, promoting both knowledge transfer and a natural regularization effect to reduce overfitting. Past studies have shown that such inductive biases lead to more robust task-generalizable models \cite{Baxter2011, Wang2022, ruder2017overviewmultitasklearningdeep, liu-etal-2019-multi}. Third, this architecture leverages the strength of convolutions in extracting local features while simultaneously using attention for the global context \cite{Wu2020LiteTW}. This design is advantageous in this case, where both local nuances and the global structure are important for learning acoustic features since acoustic information smears all time-frequency components of reverberant inputs\cite{blindEst_wang}. 

Another aspect is the training data. Many current methods rely on synthetic RIRs via RIR simulation algorithms because they can be easily generated on a large scale \cite{FullCNN_2019, ZHENG2022110901, FullCNN_2023, Suradej}. Common simulation algorithms include FDTD-based approaches such as ARD \cite{AdaptiveRecDecomp} used in \cite{FullCNN_2023}; image source-based methods such as pyroomacoustic\cite{pyroomacoustic} used in\cite{FullCNN_2023} and an RIR Generator\cite{RIR_generator} used in \cite{ZHENG2022110901}; and a stochastic approach \cite{Suradej}. However, synthetic RIRs generated from these simulation algorithms have limitations. RIRs from image source methods often simplify complex geometries and materials with sophisticated absorption and scattering coefficients, as well as wave phenomena (e.g., diffraction), due to intricate wall materials, decorations, and furniture in real rooms, to reduce computational overhead\cite{ISRA, Srivastava2022, FullCNN_2023}, although source image principle is suitable for modeling RIR. FDTD-based ARD also lacks certain real-world irregularities\cite{FullCNN_2019}, risking overfitting when large synthetic datasets are used \cite{synthetic_data_overfit}. Similarly, the stochastic method \cite{Suradej} struggles to replicate the sparse onset properties of real RIRs \cite{Traer2016}. Building on these insights, we generate synthetic data exclusively from real RIRs for blind estimation of RAPs and RGPs. To enhance adaptability to real-world conditions, we supplement our large synthetic dataset with a smaller set of real recordings. While synthetic data offer a large and easily controlled training environment to help the model quickly learn core patterns and stabilize training, real records capture nuanced characteristics of real settings. This data preparation method is given that relying solely on real-world annotated data is challenging, as such datasets are often scarce and costly to collect at a large scale. 

In addition, considering that the observed signal is restricted to a single-channel omnidirectional format that cannot sufficiently distinguish DOA without known room geometry (blind in our case)\cite{Reza2014}, we exclude DOA from our target RGPs.

\textbf{Blind estimation of occupancy level.} In contrast to RAPs and RGPs, blind occupancy level estimation has received far less attention. The few existing studies treat occupancy as a static classification problem to predict a single value rather than a time sequence\cite{occu_hvac,Chen2017AnIO}. However, in real-world settings, occupancy fluctuates over time, making instantaneous occupancy-level estimation more appropriate. Moreover, these methods ignore room acoustics, which significantly influences speech transmission. Another issue arises from their evaluation method. These methods use the F1 score rather than quantifying the distance between the predicted and actual occupancy levels (e.g., MAE), which may not provide a meaningful reflection of performance. For example, predicting 9 occupants instead of 10 is much more appropriate than predicting 2, yet both are simply ``incorrect" under a strict accuracy measure. 

Given these issues, we therefore redefine the task as instantaneous occupancy level estimation, formulating it as a classification problem at each time frame while considering room acoustics. To accomplish this, we design a dedicated model that includes an encoder and a classifier head. In addition, owing to the scarcity of publicly available data for instantaneous occupancy level estimation, we introduce a data preparation approach specifically designed to address this gap. The corresponding data generation pipeline is detailed in Section\,\ref{sec:gen_train_data}. Besides, instead of the F1 score, we use the macro-averaged MAE and MAE to evaluate the model's performance.

\section{Room Parameters}
\label{sec:rp}
\subsection{Room Acoustic Parameters}
\label{ssec:rap}
Several RAPs have been investigated and standardized \cite{toole2017sound, kuttruff2016room, CERDA200997, AcMUS, ISRA, IEC60268, ISO3382}, which are briefly introduced.

\subsubsection{Intelligibility Parameters}
\label{sssec:IntellPara}
Intelligibility parameters, including STI and $\%\rm{AL}_{\rm{cons}}$, are used to predict speech intelligibility and assess verbal comprehension in a sound field. 

\textbf{Speech intelligibility index.} The STI is employed to predict speech intelligibility and the corresponding listening difficulty in noisy surroundings. Houtgast and Steeneken initially defined the STI based on the modulation transfer function \cite{Steeneken1980}. 
The higher the STI is, the more intelligible the sound field. The STI can be calculated from the RIR, which was standardized by IEC 60286-16:2020 \cite{IEC60268}.

\textbf{Percentage articulation loss of consonants.} $\%\rm{AL}_{\rm{cons}}$ considers the proportion of misunderstood consonants, a concept originally introduced by Peutz \cite{Alcons}. Unlike the STI, which overlooks listener proficiency and linguistic factors, $\%\rm{A}_{\rm{cons}}$ extends the limitation of STI by not discounting significant information related to intelligibility and considering linguistic proficiency. 
Thus, $\%\rm{AL}_{\rm{cons}}$ emerges as a necessary complement to assess speech intelligibility comprehensively. $\%\rm{AL}_{\rm{cons}}$ can be calculated from the STI according to Farrell Becker's empirical formula \cite{FarrellBeckerformula}:
\begin{equation}
    \%\rm{AL}_{\rm{cons}} = 170.5045 \cdot e^{-5.419 \cdot \rm{STI}}.
    \label{eq:becker_alcons}
\end{equation}

\subsubsection{Reverberation Parameters}
\label{sssec:ReverbPara}
Reverberation time ($T_{60}$) and early decay time (EDT) are relevant to reverberation and quantify the subjective impression of the vivacity of a sound field. $T_{60}$ is the most essential RAP, as it characterizes the physical properties of the RACs for which the reverberation energy is distributed within -$60$ dB. The EDT represents the decay time for the initial $10$ dB to emphasize the more important contribution of early reflection to perceived reverberation. Both RAPs are derived from the RIR via Schroeder's back integration method\cite{BackInt}. We use the ISO 3382-1:2009 standard \cite{ISO3382} to derive $T_{60}$ and EDT, following the same approach adopted in several baseline studies\cite{FullCNN_2023, FullCNN_2018, CRNN_2021, CRNN_2020, Suradej, Srivastava2021BlindRP}. 

\subsubsection{Energy Parameters}
\label{sssec:EnergyPara}
Clarity ($C_{50}$ and $C_{80}$), definition ($D_{50}$), and center time ($T_s$) are the energy parameters used to measure the energy ratio of the RIR between the energy contributed from early reflections and late reverberation. They are strongly related to the impression of transparency. 
The calculation methods for these four RAPs have been standardized in ISO 3382-1:2009 \cite{ISO3382}. We use the calculation methods defined in the standard. 

\textbf{Clarity.} $C_{80}$ and $C_{50}$ express the logarithm ratios of the energy within the first $50$ ms for speech and that within the first $80$ ms for music to the remaining RIR \cite{Bradley_C80, BRADLEY1999}. 
\begin{equation}
    C_{t_e} = 10\log_{10}\Bigg(\frac{\int_{0}^{t_e}h^2(t)dt}{\int_{t_e}^{\infty}h^2(t)dt}\Bigg)
    \label{eq:C80/C50}
\end{equation}
where $t_e$ denotes $50$ or $80$ ms, respectively.

\textbf{Definition.} $D_{50}$ indicates the subjective intelligibility of speech in a room, which is defined as the ratio of the energy received within $50$ ms to the total energy of the RIR \cite{D50_ref}.
\begin{equation}
   D_{50} = \frac{\int_{0}^{50\ \rm{ms}} h^2(t) dt}{\int_{0}^{\infty} h^2(t) dt} \times 100.
   \label{eq:D50}
\end{equation}

\textbf{Center time.} $T_s$ refers to ``the center of gravity time", characterizing the balance between clarity and reverberation that is related to speech intelligibility \cite{center_time_ref}. $T_s$ is given by:
\begin{equation}
    T_s = \frac{\int_{0}^{\infty} t h^2(t) dt}{\int_{0}^{\infty} h^2(t) dt}.
\end{equation}

\subsection{Room Geometrical Parameters}
\label{ssec:rpp}
RGPs are parameters related to the geometrical characteristics of a room. These parameters encompass the room volume and the sound source distance.
\subsubsection{Room Volume}
\label{sssec:volume}
The room volume $V$ is a position-independent parameter for modeling the attributes of a room. $V$ is strongly related to the estimation of the \emph{critical distance} ($D_{c}$) \cite{kuttruff2016room}. 
$D_c$ is vital for determining whether a virtual sound source should be rendered with reverberation, thus serving as a key distance cue for the perception of reverberation by the listener \cite{kuttruff2016room, FullCNN_2019}. Jot et al. \cite{jot_augmented_2016} identified room volume as a reverberation fingerprint to characterize rooms for spatial AR rendering. $V$ also plays an important role in speech intelligibility and is used to derive the \emph{critical distance of intelligibility} ($D_{ci}$) which acts as a distance cue for perceived intelligibility \cite{vanderwerff2003what}.

\subsubsection{Sound Source Distance}
\label{sssec:distance}
The sound source distance $D$ significantly contributes to complementing the sound source localization (SSL) by integrating it with the DOA of the sound source \cite{dist_est_kushawaha, Neri_2024}. SSL is widely used in applications such as sound source separation \cite{jenrungrot_cone}, audio-oriented and navigational systems \cite{ssl_app_navi}, speech-related applications \cite{survey_ssl}, and human-robot interactions \cite{ssl_app_interact}. Furthermore, $D$ is intimately related to the perception of speech intelligibility, particularly in terms of $\%\rm{AL}_{\rm{cons}}$ \cite{vanderwerff2003what}.

\subsection{Occupancy Level}
\label{ssec:num_occu}
Detecting the instantaneous occupancy level $N$ around a listener's location in a room is highly useful for several applications. Audio- and speech-based estimation methods have attracted attention because of their nonintrusiveness, real-time estimation, and cost-effectiveness\cite{Chen2017AnIO}. The number of occupants affects reverberation \cite{Oliver2015}, thus affecting the efficacy of demand-driven hearing aid systems and speech enhancement methods. In addition, interference speeches generated by the occupants around a listener affect target signals that the listener intends to receive. Knowing the occupancy level helps control interference to achieve intelligible and clear transmission. 

In the context of smart homes, the estimated number of nearby occupants can optimize the control of demand-driven heating, ventilation, and air conditioning (HVAC) operations in the local space to significantly reduce the cost of building operations for sustainable smart buildings \cite{occu_hvac, occu_hvac_2}. In the XR and AR scenarios, the local occupancy level, as a key component of environmental information factor, is fundamental to ensuring safe interaction in real-world scenes, especially in public spaces populated by others.

\section{Preparation of Training Data}
\label{sec:gen_train_data}
This section details the preparation of training and testing data to validate our work. For the task of RAPs and RGPs estimation, we train and evaluate the model using a combination of synthetic and real datasets. For the occupancy level estimation task, we rely solely on synthetic data because of the lack of real datasets. This section has three subsections. First, we introduce the signal models used to generate synthetic data for these two tasks. Second, we describe the collections of external databases used for generating synthetic and real datasets. Third, we explain how to generate synthetic datasets for the two tasks using signal models and external databases.

\subsection{Signal Models for Synthesizing Training Data}
\subsubsection{Noisy Reverberant Signal Model}
\label{sssec:reverbNoiseSignal}
The noisy reverberant signal model is used to generate synthetic data for training the unified module in the proposed method, as shown in Fig.\,\ref{fig:block_diagram}. Details on how this model is used to synthesize the observed reverberant signal are provided in \ref{sssec:noisyReverbData}. The observed reverberant signal perceived by a listener while transmitting from a speaker within a room and subject to the influence of the background environmental noise, can be formulated as:
\begin{equation}
\label{eq:reverbNoisy}
    y_{\rm{nr}}(t) = x(t) * h(t) + n(t),
\end{equation}
where $y_{\rm{nr}}(t)$ denotes the noisy reverberant signal as perceived by the listener, $h(t)$ denotes the RIR, and $n(t)$ represents the background noise that is prevalent in the listener's local surroundings. The symbol $``*"$ denotes the convolution operation. 
$y_{\rm{nr}}$ encapsulates the RIR information that fully characterizes the RACs in the listener's local space, including the RAPs and room volume. In addition, it contains information related to the location of the sound source, such as distance. 

\subsubsection{Multitalker Reverberant Signal Model}
\label{sssec:reverbCrowd}
Due to the lack of a reverberant speech corpus for blind occupancy level estimation, we propose a signal model to simulate nearby speakers around a listener to generate synthetic data to train the occupancy module in the proposed method (Fig.\,\ref{fig:block_diagram}) to estimate the instantaneous occupancy level around the listener. 
Details of the data generation pipeline are given in Section\,\ref{sssec:noisyCrowdedData}.

Inspired by \cite{occu_hvac, Chen2017AnIO}, this signal model is formulated as:
\begin{equation}
\label{eq:reverbCrowd_init}
    y_{\rm{or}}(t) = \sum_{i=1}^{\mathcal{N}}\Big[\frac{d_0}{d_i} A_0 x_i(t) * h_i(t)\Big] + n(t),
\end{equation}
where $y_{\rm{or}}$ signifies the multitalker reverberant speech signal, $x_i(t)$ represents the speech signal originating from the $i$-th speaker proximal to the listener, and $d_i$ denotes the distance between the $i$-th speaker and the listener, which adheres to a Gaussian distribution. $\mathcal{N}$ represents the total count of occupants around an observer's location. $A_0$ represents the baseline amplitude observed at a distance of $d_0$ from the listener, i.e., the original amplitude of the speech $x_i(t)$ without attenuation. Here, $d_0$ is assumed to be equal to $1.0$, which is a proper choice considering the preferred interpersonal distance\cite{personal_distance}. $n(t)$ denotes the background noise. $h_i(t)$ denotes the RIR of each occupant with amplitude normalization. The proposed model assumes that the distance from the $i$-th speaker to the listener should exceed $d_0$, (i.e., it defines a minimal circular boundary around the speaker on the basis of the preferred interpersonal distance\cite{personal_distance}), and the amplitude corresponding to each occupant decays proportionally according to the multiplicative inverse of the associated distance\cite{occu_hvac}.

Currently, we focus on the overall effect of reverberation on overlapping speech. Given the difficulty of collecting real RIRs for every occupant, we treat early reflections as part of the direct speech signal, as they can be integrated with it under noisy conditions \cite{Bradley2003, Arweiler_Buchholz_Dau_2009,Warzybok2011InfluenceOE}.
In addition, McKenzie et al.\cite{mckenzie2023source} demonstrated that positional information is not primarily conveyed in late reverberation. Therefore, the signal model can be simplified by integrating early reflections of $h_i(t)$ into speech signals $x_i(t)$ and approximating $h_i(t)$ as a single global RIR $h(t)$. Then, Eq.\,(\ref{eq:reverbCrowd_init}) is simplified as:
\begin{equation}
\label{eq:reverbCrowd}
    y_{\rm{or}}(t) = \Big[\sum_{i=1}^{\mathcal{N}} \frac{d_0}{d_i}A_0 x_i(t)\Big] * h(t) + n(t).
\end{equation}

We acknowledge that incorporating early reflections with full position dependence would allow for a more accurate estimation of real-world scenarios. A model that explicitly accounts for early reflection will be addressed in future work.

\subsection{Dataset Collection}
\label{ssec:dataprepare}
This subsection details the collection of external datasets for the next step of dataset compilation. According to Eqs.\,(\ref{eq:reverbNoisy}) and (\ref{eq:reverbCrowd}), three datasets are required to synthesize noisy reverberant and occupancy signals: RIRs, clean speech recordings, and background noise datasets. For blind estimation of RAPs and RGPs, these datasets were used to generate synthetic data (noisy reverberant speech signals) following Eq.\,(\ref{eq:reverbNoisy}). In addition, real reverberant speech recordings were collected to improve the practical adaptability of the model. Synthetic and real datasets were combined to train the unified module, which was then evaluated separately on synthetic and real data. For the occupancy level estimation task, due to the lack of real recorded occupancy data, the occupancy module was trained and evaluated solely on synthetic data generated via Eq.\,(\ref{eq:reverbCrowd}).

\subsubsection{RIR Dataset Collection}
\label{sssec:datacollect}
 We collected real RIRs from five existing real RIR datasets to synthesize observed speech signals, including the noisy reverberant signal $y_{nr}(t)$ and the occupancy reverberant signal $y_{or}(t)$. These RIR datasets include the Arni dataset \cite{Arni}, the Motus dataset \cite{Motus}, the BUT ReverbDB \cite{BUT}, the ACE corpus \cite{ACE2016}, and the OpenAIR \cite{openair}. Each dataset comprises monochanneled and omnidirectionally recorded RIRs. All the selected datasets contain the detailed attributes in terms of room volume, covering 39 rooms with volumes ranging from 30 to $47,000$$\sim$$\rm{m}^3$, locations of sound sources and receivers or equivalent information such as sound source distances. These RIRs cover a wide range of acoustic environments, featuring various volumes and geometries of the room, varied locations of the sound source and receiver, and different sound absorption coefficients of the room surfaces. Hence, they can contain a wide spectrum of broadband RAPs and RGPs. We annotated all the labels for the collected RIRs using calculation methods in Section\,\ref{sec:rp}. While $\%AL_{\text{cons}}$ labels were derived via the empirical Eq. (\ref{eq:becker_alcons}) due to a lack of direct measurement, its limited applicability motivated us to predict $\%AL_{\text{cons}}$ end-to-end, aiming for adaptability to potential future scenarios where $\%AL_{\text{cons}}$ might be obtained directly, not just via the STI. We resampled all RIRs to $16$ kHz. 

\subsubsection{Clean Speech Dataset Collection}
\label{sssec:speechdatacollect}
We use the LibriSpeech corpus \cite{librispeech} to sample clean speech signals when synthesizing observed reverberant signals. Specifically, we select a 360-hour clean subset. This subset is composed of more than 100,000 unique clips articulated by 921 speakers with completely distinct linguistic contents. The deployment of this dataset ensures a broad spectrum of diverse speech signals, enhancing the robustness and generalizability of our synthesized signals.

\subsubsection{Background Noise Dataset Collection}
\label{sssec:noisedatacollect}
To replicate the environmental background noise encountered in real-world scenarios, instead of using synthetic white Gaussian noise, we use actual noise samples from real-world daily life circumstances. We integrate noise signals from the DEMAND \cite{DEMAND} and BUT\cite{BUT} noise datasets, both of which are collected in real-world environments of daily life and resampled at $16$ kHz.

\subsubsection{Real Reverberant Speech Recordings Collection}
For the blind estimation task of RAPs and RGPs, we mixed a smaller number of real reverberant speech recordings with large-scale synthetic reverberant data to train and evaluate the model. In total, we collected 11,160 samples from the BUT Retransmission dataset \cite{BUT}, where clean speech is transmitted in real rooms to obtain real recordings.

\subsection{Synthetic Dataset Generation}
\label{ssec:dataaug}
\subsubsection{Synthesis Pipeline of Noisy Reverberant Speech Signals}
\label{sssec:noisyReverbData}
In the collected real RIR dataset with labels of RAPs and RGPs, we further employ a data augmentation strategy. The strategy involves \emph{data upsampling} and \emph{downsampling} techniques to modulate the distribution of labels. For data upsampling, we repeat some RIRs in which those labels appear less frequently to convolve with more clean speech recordings. Because each clean speech recording is distinct, we still guarantee that the generated examples have enough diversity after data upsampling. Conversely, for data downsampling, we skip some RIRs that have abundant labels. The degrees of upsampling and downsampling are calibrated based on the relative rarity of labels. For example, collected real RIR datasets often include many labeled examples with smaller volumes (e.g., $<400\sim m^3$) but very few with larger volumes (e.g., $7000-10000\sim m^3$), resulting in an imbalanced distribution of room volume labels. To address this imbalance, we downsampled RIRs associated with more prevalent room volume labels while concurrently upsampling RIRs corresponding to the rarer volume labels. This data augmentation strategy is open-sourced in the data preprocessing pipeline\footnote{\url{https://github.com/Alizeded/BERP/blob/main/notebooks/dataset_preprocess.ipynb}}. Afterward, a comprehensive collection of $47,430$ real RIRs is successfully compiled. This RIR dataset contains a wide range of RIRs, for which the corresponding $T_{60}$ spans from $0.19$ to $7.88$$\sim$s. The mean reverberation time ($T_{60}$) across this dataset is approximately $1.73$$\sim$s.

We randomly sample 47,430 clips from the LibriSpeech corpus as clean speech signals, regardless of the speaker information and linguistic content they contain. In parallel, noise signals are randomly sampled, following an independent and identically distributed (I.I.D.) pattern, from the DEMAND and BUT datasets. Then, according to Eq.\,(\ref{eq:reverbNoisy}), we synthesize the noisy reverberant speech signals via RIR, clean speech, and noise signals, for a total of 261 hours. To increase the robustness and efficacy of the model in various noisy environments, the signal-to-noise ratio (SNR) between reverberant and noisy signals is uniformly varied by adjusting the SNR at five different levels, ranging from 0 to 20$\sim$dB in 5$\sim$dB increments, including a scenario without noise (Inf). Given the uniqueness of each clip, we guarantee that every synthesized speech signal maintains its individuality in terms of both its waveform and linguistic content, further augmenting the diversity and richness of the synthetic dataset.

\subsubsection{Synthesis Pipeline of Multitalker Reverberant Speech Signals}
\label{sssec:noisyCrowdedData}
First, we applied voice activity detection to the LibriSpeech corpus to segment and annotate timestamps corresponding to speech and silence segments. This process underlies annotations of the synthesized multitalker signals. 

\begin{figure}[h]
\vspace{-0.2em}
    \centering
    \includegraphics[width=0.33\textwidth]{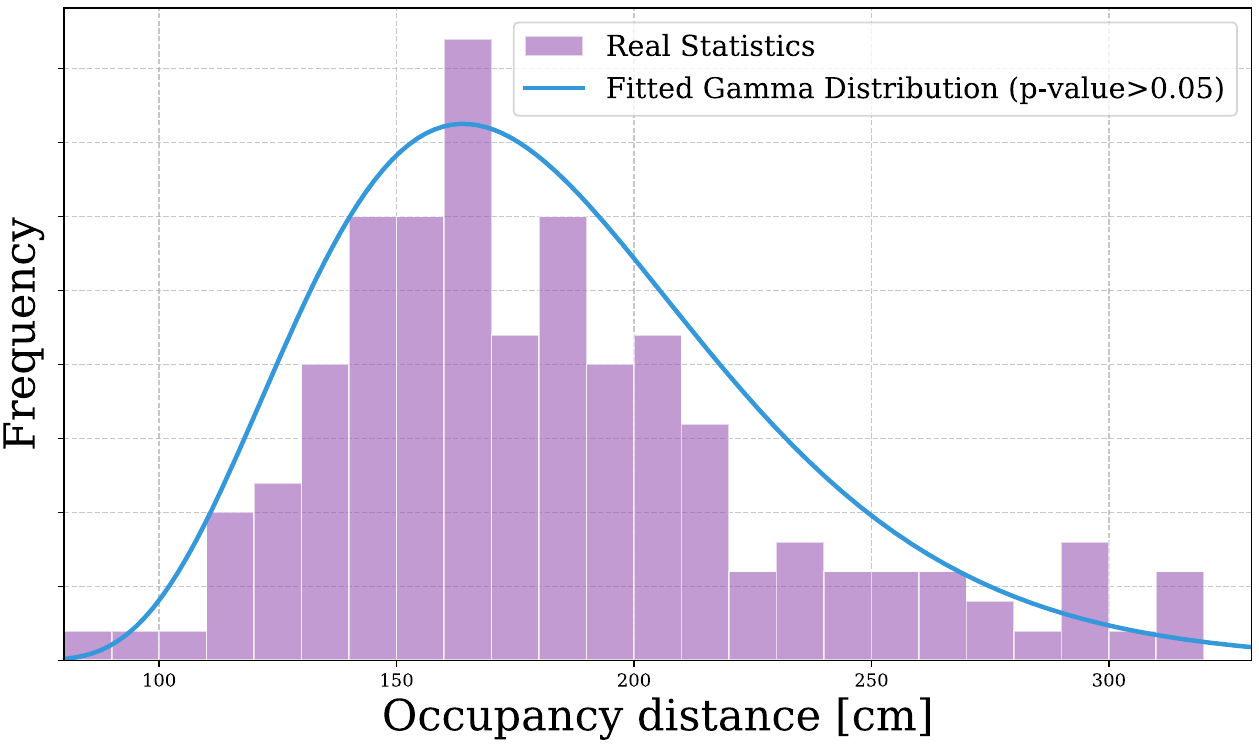}
    \vspace{-0.5em}
    \caption{Histogram of occupancy distances}
    \label{fig:hist_occu_dist}
    \vspace{-1.1em}
\end{figure}

Next, we model each occupant's distance $d_i$ from the listener via a Gamma distribution $d_i \sim \Gamma(k, \theta)$, where $k$ and $\theta$ are the shape and scale parameters, respectively. To validate this setting, we fit real occupancy-distance data from \cite{STARSS23} (see Fig.\,\ref{fig:hist_occu_dist}) and used a chi-square goodness-of-fit test, resulting in a p-value above 0.05. This finding indicates that a Gamma distribution effectively models the occupancy distance. Consequently, we set $k = 6.18$ and $\theta = 18.66$. Furthermore, we limit the occupancy  distance $d_i$ up to 6$\sim$m, according to the considerations discussed below. The normal speech level at 1$\sim$m (56--58$\sim$dB\cite{normal_speech_spl}) attenuates to 34--36$\sim$dB at 6$\sim$m (5$\sim$m distance)\footnote{Calculated based on ISO 9613-2:2024\cite{ISO9613} without considering reflections in a typical room setting at 23$^{\circ}\text{C}$ and $50\%$ humidity \cite{WOLKOFF2021113709}.}. In addition, given the early reflection effect for speech intelligibility with a 4$\sim$dB improvement at a distance of 5$\sim$m in noisy conditions\cite{Warzybok2011InfluenceOE}, the speech sound level at 6$\sim$m is 38--42$\sim$dB. Such levels are easily masked by background noise in most daily scenarios, so we treat occupant speech beyond this distance as part of background noise. 

\begin{figure}[h]
\vspace{-0.1em}
    \centering
    \includegraphics[width=0.33\textwidth]{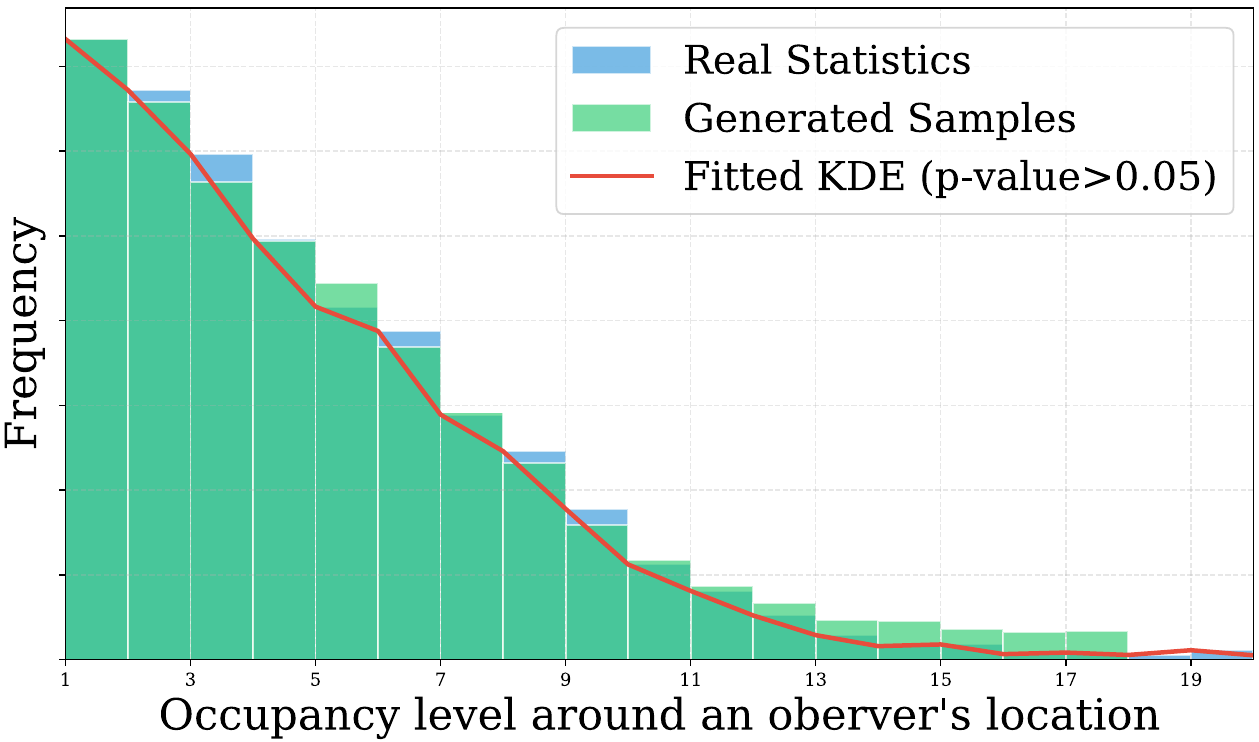}
    \vspace{-0.3em}
    \caption{Histogram of instantaneous occupancy levels.}
    \label{fig:hist_occu}
    \vspace{-1.4em}
\end{figure}

Meanwhile, we used statistics on the real-world occupancy level and employed kernel density estimation (KDE) \cite{Chen01012017} to determine the occupants (i.e., the number of individuals) for each synthesized overlapping sample. Note that this number represents the speech numbers used to synthesize overlapping speech rather than the instantaneous occupancy level at each time frame. After analyzing a large-scale real occupancy level statistics from \cite{Tekler2022}, we determined that traditional parametric distributions (e.g., discretized Gamma or negative binomial) do not adequately capture the complexity of the observed patterns. Therefore, we instead use a nonparametric approach to model the occupancy level distribution. The chi-square test (p-value$>$0.05) confirms that KDE effectively represents the underlying distribution (see Fig.\,\ref{fig:hist_occu}). Then, we use this validated KDE model to generate occupancy levels that closely mirror real-world settings. Figure\,\ref{fig:hist_occu} depicts the resulting occupancy distribution in the compiled synthetic dataset. Here, we set the maximum $N$ to 17 within the 6$\sim$m (max $d_i$) radius, on the basis of the rationale presented below. First, given that the area of this radius is approximately 110.0$\sim$$\rm{m}^2$ and that the preferred social distance is $1.4$$\sim$m (6.2$\sim$$\rm{m}^2$) per person \cite{personal_distance}, 17 is a reasonable capacity in this area. Second, given the difficulty of detecting occupancy levels from overlapping speech, 17 is the maximum occupancy level that we can now detect. Notably, these considerations are a plausible practical guide rather than a definitive scientific conclusion. Real environments may exceed this threshold, and more studies are needed to validate these findings under various acoustic conditions.

Afterward, we aligned each occupancy level $N$ with RIR data according to the room volume, following the real-world principle that larger rooms generally accommodate more occupants. Specifically, we classified the room volume into five ranges delimited at 400, 2000, 4000, and 6000$\sim$$\rm{m}^3$. Each range was assigned an upper bound for $N$. For example, rooms under 400$\sim$$\rm{m}^3$ have an upper bound of 7, and we randomly choose one value from 1 to 7, while volumes between 400 and 2000$\sim$$\rm{m}^3$ have an upper bound of 12, and so on. However, to account for less common cases in which smaller rooms may hold more people, we did not strictly limit $N$ to a subset of the data, thereby allowing some smaller rooms to exceed these thresholds. The implements can be found in our GitHub repository.
\begin{figure}[htbp]
\vspace{-0.9em}
    \centering
    \includegraphics[width=0.35\textwidth]{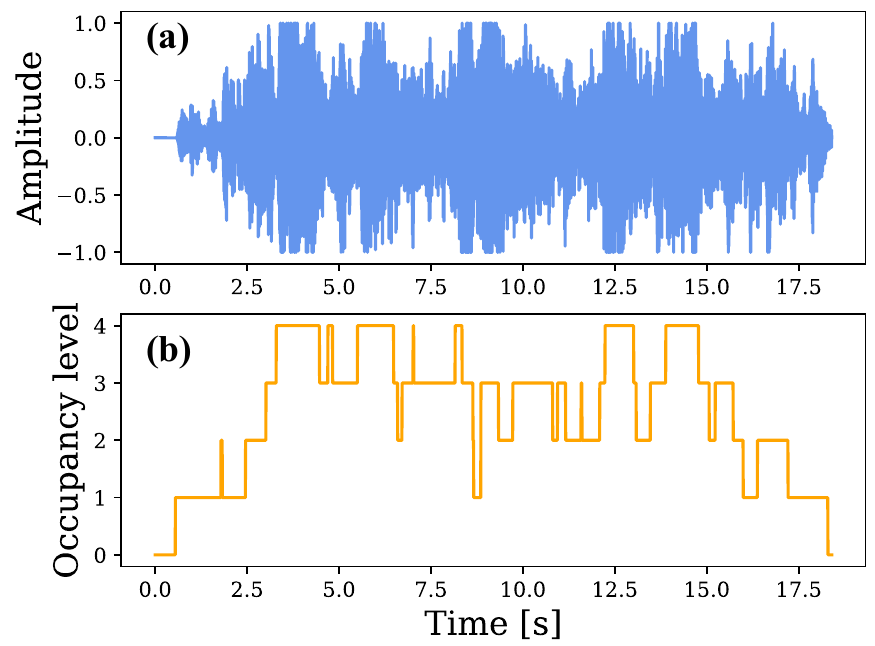}
    \vspace{-0.5em}
    \caption{An example of a multitalker reverberant speech signal. (a) The occupancy reverberant signal. (b) The corresponding instantaneous occupancy level. (smoothed in frame 500 ms)}
    \label{fig:illustration_crowdspeech}
    \vspace{-0.5em}
\end{figure}

Finally, via Eq.\,(\ref{eq:reverbCrowd}), we synthesize occupancy reverberant speech by overlapping speech signals uniformly sampled from the LibriSpeech corpus, convolving with real RIRs, adding background noise, and aligning with annotated speech and silence segments. The background noise is randomly sampled from the compiled noise dataset mentioned in Section\,\ref{sssec:noisedatacollect}. The noise SNR is uniformly adjusted across six levels, ranging from $25$ dB to $50$ dB in the $5$ dB increments, to align with of that background noise originates away from the 6-meter circular area. The initiation index for each overlapping speech signal is determined on the basis of an I.I.D. pattern. Additionally, to replicate the local room acoustics, the room volumes in the RIRs are precisely matched with their corresponding RIRs, thereby ensuring a realistic acoustic environment. Furthermore, when convolving the RIRs, we preprocessed the RIRs to discard position-dependent information. Specifically, we filtered the direct sound and early reflection (initial $50$ $\text{ms}$) of the RIR (see Section\,\ref{sssec:reverbCrowd} for more details). We also normalize the amplitude of $h_i(t)$ to eliminate the cues related to the sound source distances from the RIR amplitudes. These steps discard position-dependent information to the greatest extent feasible, ensuring that the RIR used in this signal model represents only late reverberation-related acoustical characteristics. 
 We obtain a dataset comprising 47,430 samples of multitalker reverberant speech signals, ranging from 10 to 25$\sim$s, for a total of 222 hours of audio. Figure\,\ref{fig:illustration_crowdspeech} shows an example of a multitalker reverberant speech signal and its corresponding instantaneous occupancy level.

\section{Proposed Approach}
\label{sec:proposed}
\noindent\textbf{Overview: BERP.} Figure\,\ref{fig:block_diagram} shows the proposed BERP approach. It comprises two networks, each trained independently for its specific task. One (the unified module) processes a noisy reverberant speech signal $y_{\mathrm{nr}}(t)$ to estimate RAPs and RGPs, corresponding to a scenario in which a primary speaker with background noise (e.g., distant speech, clapping, etc.) within the room. The other (the occupancy module) processes a multitalker reverberant speech signal $y_{\mathrm{or}}(t)$ to estimate instantaneous occupancy levels, assuming that the room contains multiple overlapping speakers against background noise. Note that before being fed into the RFE, $y_{nr}(t)$ and $y_{or}(t)$ are transformed into time-frequency representations via featurizer.

\vspace{-0.5em}
\subsection{Problem Formulation}
\textbf{Unified Module.} As mentioned in Section\,\ref{sec:intro}, we formulate the blind estimation of RAPs and RGPs as a multitask learning problem. We can yield the following objective function:
\begin{equation}
    \underset{\boldsymbol{\theta^{sh}}, \boldsymbol{\theta_1^t}, \cdots, \boldsymbol{\theta_n^t}}{\min} \sum_{i=1}^{n}\mathcal{L}^{\text{unf}}_{\gamma_i}(f_i^{\gamma}(y_{nr}(t);\boldsymbol{\theta}^{sh},\boldsymbol{\theta}_i^{t}), \gamma_{i}).
    \label{eq:unf_obj_func}
\end{equation}
$f_i^{\gamma}: \mathbb{R}^{1\times L} \rightarrow \mathbb{R}^{1}$ is the trainable mapping function between the observed signal $y_{nr}(t)\in 1 \times L$ with length $L$ and the estimated room parameter $\hat{\gamma}\in \mathbb{R}^{1}$, i.e., $f_i^{\gamma}(y_{nr}(t);\boldsymbol{\theta}^{sh},\boldsymbol{\theta}_i^{t}) = \hat{\gamma}_{i}$. $\mathcal{L}_{\gamma_i}^{\text{unf}}$ is the corresponding loss between the estimated room parameter $
\hat{\gamma_{i}}$ and the ground-truth room parameter $\gamma_{i}$. $\boldsymbol{\theta}^{sh}$ represents the shared parameters of the neural network model, whereas $\boldsymbol{\theta}_i^{t}$ represents the task-specific parameters for each $\gamma_{i}$. $n$ is total number of room parameters. This objective function optimizes the parameters $\boldsymbol{\theta}^{sh}$ and $\boldsymbol{\theta}^{t}$ to obtain Pareto optimality\cite{Wang2022}. We use a shared encoder RFE to represent $\boldsymbol{\theta}^{sh}$ and multiple task-specific PPs as $\boldsymbol{\theta}^{t}$ for each room parameter. 

\textbf{Occupancy Module.} We formulate the following objective function to minimize the difference between the predicted and ground-truth probabilities of occupancy levels:
\begin{equation}
    \underset{\boldsymbol{\theta}^{o}}{\min} -\log p_{\boldsymbol{\theta}^{o}}(N|f^{o}(y_{or}(t);\boldsymbol{\theta}^{o})).
    \label{eq:occu_obj_func}
\end{equation}
$f^{o}: \mathbb{R}^{1 \times L} \rightarrow \mathbb{R}^{1 \times T}$ is the trainable mapping function between the observed signal $y_{or}(t)\in \mathbb{R}^{1 \times L}$ and the estimated instantaneous occupancy level $\hat{N}\in \mathbb{R}^{1 \times T}$, i.e., $f^{o}(y_{or}(t);\boldsymbol{\theta}^{o}) = \hat{N}$. $\boldsymbol{\theta}^{o}$ represents the parameters of the occupancy module that need to be optimized. $\boldsymbol{\theta}^{o}$ consists of an RFE and an FC layer.

The architectural details of the RFE and the PP are discussed in Section \ref{ssec:framework_arch}.

\subsection{Model Architecture}
\label{ssec:framework_arch}
\subsubsection{Featurizer}
\label{sssec:featurizer}
We use four types of featurizers to transform the observed input signals, including \emph{spectrogram}, \emph{Gammatonegram}, \emph{MFCC}, and \emph{mel spectrogram}. The Gammatonegram emphasizes the importance of low-frequency sections while a signal propagates within a room \cite{FullCNN_2018}, whereas MFCC characterizes the shape of the spectral envelope of a reverberant signal, which is closely related to the MTF of room acoustics \cite{schroeder1981}. The mel spectrogram mimics human subjective perceptions of RACs. To clarify the potential benefits of auditory-perceptually based time-frequency representations for these blind estimation tasks, we also use spectrograms. For an input signal $y(t)\in \mathbb{R}^{1 \times L}$  of length $L$ (i.e., $y_{nr}(t)$ and $y_{or}(t)$), the featurizer converts it into an $F\times T$ time-frequency representation, before passing it to the RFE for further processing. Here, $T$ is the time frame length, and $F$ is the number of frequency bins.

\subsubsection{Room Feature Encoder}
Inspired by \cite{Wu2020LiteTW, conformer}, we combine convolutional layers and the attention mechanism to capture local and global contexts together in time-frequency representations of input speech signals. As Wang\cite{blindEst_wang} reported that acoustic information is spread throughout the entire time-frequency component of a reverberant signal, and paying attention locally and globally together in a time-frequency map of a reverberant speech with a complicated waveform is well suited for learning more representative and robust features. 

The RFE is structured into eight blocks, each of which consists of four components. It incorporates a half-residual feedforward network, a multiheaded self attention, a convolutional network, and another half-residual feedforward network \cite{conformer}. Figure\,\ref{fig:block_diagram} illustrates the architecture of the RFE. When a spectrogram-variant feature $S\in \mathbb{R}^{F \times T}$ is fed into the RFE, the RFE first maps it to a tensor $\mathbcal{x}_{\text{RFE}} \in \mathbb{R}^{C \times T}$ via a fully connected layer (FC). Here, $C$ is the number of embedding dimensions. We set $C$ to 512.

The forward pass of RFE at block $i$ can be expressed as:
\begin{equation}
    \mathbcal{x}_{i, \text{RFE}}^{\ddagger, \mathbb{R}^{C \times T}} = \mathbcal{x}_{i, \text{RFE}}^{\mathbb{R}^{C \times T}} + \frac{1}{2} \cdot \textbf{FFN}(\mathbcal{x}_{i, \text{RFE}}^{\mathbb{R}^{C \times T}}),
\end{equation}
\begin{equation}
    \mathbcal{x}_{i, \text{RFE}}^{\ddagger\ddagger, \mathbb{R}^{C \times T}} = \mathbcal{x}_{i, \text{RFE}}^{\ddagger, \mathbb{R}^{C \times T}} + \text{LayerNorm}[\textbf{MHSA}_{\text{xPos}}(\mathbcal{x}_{i, \text{RFE}}^{\ddagger, \mathbb{R}^{C \times T}})],
\end{equation}
\begin{equation}
    \mathbcal{x}_{i, \text{RFE}}^{\ddagger\ddagger\ddagger, \mathbb{R}^{C \times T}} =\mathbcal{x}_{i, \text{RFE}}^{\ddagger\ddagger, \mathbb{R}^{C \times T}} + \textbf{Conv}(\mathbcal{x}_{i, \text{RFE}}^{\ddagger\ddagger, \mathbb{R}^{C \times T}}),
\end{equation}
\begin{equation}
    \mathbcal{y}_{i, \text{RFE}}^{\mathbb{R}^{C \times T}} = \text{LayerNorm}\Big[\mathbcal{x}_{i, \text{RFE}}^{\ddagger\ddagger\ddagger, \mathbb{R}^{C \times T}} + \frac{1}{2} \cdot \textbf{FFN}(\mathbcal{x}_{i, \text{RFE}}^{\ddagger\ddagger\ddagger, \mathbb{R}^{C \times T}})\Big],
\end{equation}
where $\mathbcal{x}_{\text{RFE}}$ and $\mathbcal{y}_{\text{RFE}}$ are the input and output of one block, respectively. $\textbf{FFN}$, $\textbf{MHSA}_{\text{xPos}}$, and $\textbf{Conv}$ denote the feedforward network, multiheaded self-attention, and convolutional network, respectively. LayerNorm represents the layer normalization operation.

\textbf{Feedforward network.} The \textbf{FFN} consists of two FC layers with prenorm layer normalization and a dropout layer (dropout rate: $0.1$) as the final layer. The first FC layer is an intermediate layer with 2048 dimensions, whereas the second is the output layer with 512 embedding dimensions. The architecture is detailed in Fig.\,\ref{fig:block_diagram}, functioning as $f_{\textbf{FFN}}: \mathbb{R}^{512 \times T} \rightarrow \mathbb{R}^{2048 \times T} \rightarrow \mathbb{R}^{512 \times T}$. 
The activation function used after the intermediate layer is the swish function\cite{swish}, defined as
\begin{equation}
    \text{swish}(x) = x \cdot \text{sigmoid}(x) = \frac{1}{1+e^{-x}}.
\end{equation}

\textbf{Multiheaded self-attention.} We employ the \textbf{MHSA} with the extrapolatable relative positional encoding (xPos) along the time frames $T$. xPos-based self-attention has been shown to better capture long-term dependencies than vanilla or rotary positional encoding methods \cite{xpos}. This enhanced ability to handle global contexts is crucial in our case because the acoustical information in the reverberant signal is spread across the overall time-frequency domain \cite{blindEst_wang}. Using xPos, our model can handle these globally dispersed features more effectively, thus improving the performance of the attention mechanism. Moreover, xPos encoding has been empirically validated to augment the stabilization and robustness of the self-attention mechanism, particularly for long-sequence inputs\cite{xpos}, such as audio signals.

The xPos-based self attention can be formulated as follows\cite{xpos}:
\begin{equation}
    \textbf{MHSA}_{\text{xPos}}(\mathbcal{x})= \text{Softmax}\Bigg(\frac{\boldsymbol{Q}_{\mathbcal{x}, \rm{xPos}} \boldsymbol{K}_{\mathbcal{x}, \rm{xPos}}^\text{T}}{\sqrt{C_h}} \boldsymbol{M}\Bigg)\boldsymbol{V}_{\mathbcal{x}},
\end{equation}
where $\boldsymbol{Q}_{\mathbcal{x}, \rm{xPos}}:\mathbb{R}^{H \times C_h \times T} = (\textbf{W}_q \mathbcal{C}+\mathfrak{R}_{\boldsymbol{Q}}\mathbcal{S})\mathbcal{T}$, $\boldsymbol{K}_{\mathbcal{x}, \rm{xPos}}:\mathbb{R}^{H \times C_h \times T} = (\textbf{W}_k \mathbcal{C}+\mathfrak{R}_{\boldsymbol{K}}\mathbcal{S})\mathbcal{T}^{-1}$, and $\boldsymbol{V}_{\mathbcal{x}}:\mathbb{R}^{H \times C_h \times T} = \textbf{W}_v \mathbcal{x}$. $H$ is the number of heads of attention mechanism. Here, we set $H$ as 8. $C_h$ is the head dimensions, which is 64. $\mathbcal{C}:\mathbb{R}^{C_h \times T}$ is equal to $\cos(m\vartheta_i)$ and $\mathbcal{S}:\mathbb{R}^{C_h \times T}$ is equal to $\sin(m\vartheta_i)$, which are the cosine and sine positions in the embedding dimension $i$ and the time frame $m$, respectively. $\mathfrak{R}$ corresponds to the rotary matrix of $\boldsymbol{Q}$ and $\boldsymbol{K}$, which is $\mathfrak{R}({\boldsymbol{\eta}}) = [-\eta_1, \eta_0, -\eta_3, \eta_2, ...]$. ``T" denotes transposition. $\mathbcal{T} = \varsigma_{m, i}$. The $\varsigma_i$ is given by:
\begin{equation}
    \varsigma_i = \frac{i/\frac{\mathbcal{D}_h}{2}+\beta}{1+\beta},
\end{equation}
where $\beta$ is optimally set to 512,
and $\vartheta_i = 10000^{-2i/\mathbcal{D}_h}$. $\textbf{W}_q$, $\textbf{W}_k$, $\textbf{W}_v$ and $\boldsymbol{M}$ are trainable weighting matrices of the query, key, and value of the attention mechanism, respectively. Specifically, the tensor that passes through the \textbf{MHSA} is as follows: $\mathbb{R}^{C \times T} \rightarrow  \mathbb{R}^{H \times C_h \times T} \rightarrow \mathbb{R}^{C \times T}$.

\textbf{Convolutional network.} The convolutional network functions to capture the local features and strengthen the temporal causality of the feature representation. This module leverages prenorm residual connections with gating mechanisms to distill the important features. It consists of a $1 \times 1$ pointwise 1-D convolutional layer with a gated linear unit (GLU) \cite{GLU}, a $31 \times 1$ 1-D depthwise convolutional layer with swish activation function, and another $1 \times 1$ 1-D pointwise convolutional layer. Assuming that the input is $\mathbcal{x}\in \mathbb{R}^{C \times T}$, the forward pass of the convolutional network can be formulated as:
\begin{subequations}
    \begin{eqnarray}
     \mathbcal{x}^{\prime, \mathbb{R}^{C \times T}} & = & \text{LayerNorm}[\text{GLU}(\text{conv}^{pw}(\mathbcal{x}^{\mathbb{R}^{C \times T}}))], \\
     \mathbcal{x}^{\prime\prime, \mathbb{R}^{C \times T}} & = & \text{conv}^{pw}[\text{swish}(\text{conv}^{dw}(\mathbcal{x}^{\prime, \mathbb{R}^{C \times T}}))],
\end{eqnarray}
\end{subequations}
where $\text{conv}^{pw}$ and $\text{conv}^{dw}$ denote as the pointwise and depthwise convolutional layers, respectively. GLU denotes the gated linear unit operation. 

\subsubsection{Parametric Predictor}
\label{sssec:param_predictor}
As shown in Fig.\,\ref{fig:block_diagram}, the PP consists of four convolutional layers with ReLU activation functions, a dropout layer and an FC layer, which together compose $\boldsymbol{\theta}^{t}$ to predict the respective RAP and RGP from the shared latent space \(\mathbcal{y}_{\text{RFE}} \in \mathbb{R}^{C \times T}\) produced by the RFE. The first two and the final layers are $3 \times 1$ 1-D convolutional layers, whereas the third is a $5 \times 3$ 1-D convolutional layer. Each convolutional layer has 384 embedding dimensions, enabling the learning of task-specific features in a high-dimensional space. The convolution operation is applied along the time axis in the same direction, ensuring that temporal causality is preserved time-frame-by-time-frame. The final FC layer reduces to a single dimension, outputting the room parameter $\boldsymbol{\hat{\gamma}} \in \mathbb{R}^{1 \times T_{PP}} $ for each time frame. In essence, the PP predicts a straight line along time frames $T_{PP}$ to handle various input lengths without alignment. 

We show the mapping function of the PP from learned room features $\mathbcal{y}_{\text{RFE}}^{\mathbb{R}^{T \times C}}$ to a dedicated room parameter $\boldsymbol{\hat{\gamma}}^{\mathbb{R}^{1 \times T_{PP}}}$ as follows:
\begin{subequations}
    \begin{eqnarray}
    \mathbcal{y}_{\gamma}^{\mathbb{R}^{384 \times T_{PP}}} & = & \underbrace{\text{ReLU}[\text{conv}^{PP}(\mathbcal{y}_{\text{RFE}}^{\mathbb{R}^{C \times T}})]}_{4 \times}, \\
    \boldsymbol{\hat{\gamma}}^{\mathbb{R}^{1 \times T_{PP}}} & = & \text{FC}(\mathbcal{y}_{\gamma}^{\mathbb{R}^{384 \times T_{PP}}}).
\end{eqnarray}
\end{subequations}
$\text{conv}^{PP}$ are the convolutional layers used in the PP. During inference, we sum up and average the values over the time axis to obtain a single predicted room parameter $\hat{\gamma} \in {\mathbb{R}^{1}}$. The architecture of the PP is illustrated in Fig.\,\ref{fig:block_diagram}. 

\subsubsection{Classifier for Occupancy Level Estimation}
\label{sssec:occu_estimator}
An FC layer with a Softmax activation function is used as a classifier head to predict the instantaneous occupancy level $\hat{N} \in \mathbb{R}^{1 \times T}$ as a time sequence from the learned room feature $\mathbcal{y}_{\text{RFE}}^{\mathbb{R}^{C \times T}}$. The resolution of the estimation process is 1$\sim$s per time frame.

\subsection{Loss Function}
\label{ssec:lossfunc}
\subsubsection{Loss for the Unified Module}
Huber loss \cite{Huber_loss} is employed as $\mathcal{L}^{\text{unf}}_{\gamma}$ for each dedicated room parameter $\gamma$ in the unified module, which is given by:
\begin{equation}
\label{eq:predictor_loss}
    \mathcal{L}_{\gamma}^{\text{unf}}(\gamma, \hat{\boldsymbol{\gamma}}) =
    \begin{cases}
        \frac{1}{2} \sum_{t=1}^{T_{PP}} (\gamma - \hat{\gamma}_{t})^2, \hspace{19pt} \lvert \gamma - \hat{\gamma}_{t} \rvert \leq \delta \\
        \delta \sum_{t=1}^{T_{PP}} \lvert \gamma - \hat{\gamma}_{t} \rvert - \frac{1}{2}\delta^2, \hspace{4pt} \lvert \gamma - \hat{\gamma}_{t} \rvert > \delta
    \end{cases}
\end{equation}
    where $\gamma \in \mathbb{R}^{1}$ is a ground-truth room parameter and $\hat{\gamma}_t \in \mathbb{R}^{1}$ is the estimated room parameter at each time frame $t$. $\delta$ is set to 1. 
    $T_{PP}$ is the output time frame from the PP. Huber loss has dual sensitivities of minimum-variance estimation and robustness against outliers. It avoids the convergence problem of $\mathcal{L}$1 loss on a small scale and contributes to preventing exploding gradients by clipping gradients exceeding $\delta$ \cite{loss_survey2023}.  

\subsubsection{Loss for the Occupancy Module}
Specifically, we use cross-entropy (CE) to measure the difference between the probabilities of the estimated and ground-truth occupancy levels (the term $-\log p_{\boldsymbol{\theta}^{o}}$ in Eq.\,(\ref{eq:occu_obj_func})), reflecting the multiclass nature of the task. The CE loss function is determined as follows: 
\begin{equation}
\label{eq:occu_loss}
    \mathcal{L}^{o} =  -\log p_{\boldsymbol{\theta}^{o}}(N|\hat{N}) = -\sum_{c=0}^{\mathcal{C}} N_{c} \log\big(\hat{p}^{o}_{c}),
\end{equation}
\vspace{-0.5em}
\begin{equation}
    \hat{N}_{c} = \text{argmax}(\hat{p}^{o}_{c}) \notag 
\end{equation}
where $N_c \in \mathbb{R}^{1 \times T_{c}}$ is the true label of the occupancy level class $c$, represented as 1 for the correct class and 0 for otherwise. $T_{c}$ denotes the respective time frame for each class $c$. $\hat{p}^{o}_{c} \in \mathbb{R}^{1 \times T_{c}}$ is the predicted probability of the class $c$ being output from the occupancy module. Here, $\mathcal{C}$ is set to 17 according to Section\,\ref{sssec:noisyCrowdedData}.

\begin{table*}[!t]
    \centering
    \caption{Comparison among BERP variants of four types of featurizers, including MFCC, mel spectrogram, Gammatonegram and spectrogram. \textbf{Bold} represents the best.}
    \begin{tabular}{ccccccccccccc}
    \specialrule{.2em}{.1em}{.1em}
         & & STI & $\%\text{AL}_{\text{cons}}$ & $T_{60}$ & EDT & $C_{80}$ & $C_{50}$ & $D_{50}$ & $T_s$ & $V$ & $D$ \\
         & Featurizer & & & [$s$] & [$s$] & [dB] & [dB] & [\%] & [$s$] & [$\log_{10}(m^3)$] & [$m$] \\
         \specialrule{.1em}{.05em}{.05em} 
         \multirow{4}{*}{MAE $\downarrow$}
         & MFCC & \textbf{0.0049} & \textbf{0.1523} & \textbf{0.0162} & \textbf{0.0198} & \textbf{0.2256} & \textbf{0.2231} & 1.041 & \textbf{0.0025} & 0.0111 & 0.1864 \\
         & Mel spectrogram & 0.0055 & 0.1716 & 0.0178 & 0.0223 & 0.2683 & 0.2800 & 1.120 & 0.0026 & \textbf{0.0096} & 0.2402 \\
         & Gammatonegram & 0.0082 & 0.2940 & 0.0289 & 0.0407 & 0.4833 & 0.5040 & 2.069 & 0.0049 & 0.0210 & 0.6160 \\
         & Spectrogram & 0.0054 & 0.2100 & 0.0198 & 0.0251 & 0.2396 & 0.2394 & \textbf{0.9374} & 0.0028 & 0.0167 & \textbf{0.1791} \\
         \hline 
         \multirow{4}{*}{PCC $\uparrow$}
         & MFCC & 0.9960 & \textbf{0.9964} & 0.9973 & \textbf{0.9979} & \textbf{0.9964} & \textbf{0.9950} & \textbf{0.9947} & \textbf{0.9955} & 0.9819 & \textbf{0.9716} \\
         & Mel spectrogram & \textbf{0.9968} & \textbf{0.9964} & \textbf{0.9977} & 0.9966 & 0.9932 & 0.9919 & 0.9935 & 0.9954 & \textbf{0.9867} & 0.9695 \\
         & Gammatonegram & 0.9932 & 0.9956 & 0.9961 & 0.9936 & 0.9843 & 0.9796 & 0.9838 & 0.9897 & 0.9754 & 0.8647 \\
         & Spectrogram & 0.9940 & 0.9925 & 0.9974 & 0.9953 & 0.9947 & 0.9917 & 0.9924 & 0.9923 & 0.9581 & 0.9715 \\
    \specialrule{.2em}{.1em}{.1em}
    \end{tabular}
    \label{tab:feat_compare}
\end{table*}
\begin{table*}[!t]
  \centering
    \caption{MAE and PCC comparisons among the proposed BERP and baselines for RAPs and RGPs on \emph{real} and \emph{synthetic} data. After the comparative experiment in Table\,\ref{tab:feat_compare} of four types of featurizers, the MFCC-based BERP variant is chosen for comparison with the baselines. $\textbf{Bold}$ indicates the best result, and ``-" indicates no data available or no measurement performed.}
    \begin{tabular}{ccccccccccccc}
    \specialrule{.2em}{.1em}{.1em} 
        & & STI & $\%\text{AL}_{\text{cons}}$ & $T_{60}$ & EDT & $C_{80}$ & $C_{50}$ & $D_{50}$ & $T_s$ & $V$ & $D$ \\
        & Model & & & [$s$] & [$s$] & [dB] & [dB] & [\%] & [$s$] & [$\log_{10}(m^3)$] & [$m$] \\
       \specialrule{.1em}{.05em}{.05em} 
       & ($\emph{\textbf{Real}}$) \\
       \multirow{13}{*}{MAE $\downarrow$}
       & Full-CNN\cite{FullCNN_2023} & - & - & 0.6945 & - & - & - & - & - & 0.2804 & -\\
       & CRNN\cite{CRNN_2021} & 0.0514 & - & 0.1817 & - & 2.123 & 2.021 & - & - & -\\
       & CNN-MLP\cite{Srivastava2021BlindRP} & - & - & 0.1156 & - & & - & - &- & 0.1513 & - \\
       & TAE-CNN\cite{Suradej} & 0.1778 & - & 0.8107 & 0.8698 & 7.139 & - & 31.08 & 0.0659 & - & - \\
       & RE-NET\cite{ZHENG2022110901} & - & - & 0.4626 & - & - & - & - & - & - & - \\
       & BERP (Ours) & \textbf{0.0049} & \textbf{0.1523} & \textbf{0.0162} & \textbf{0.0198} & \textbf{0.2256} & \textbf{0.2231} & \textbf{1.041} & \textbf{0.0025} & \textbf{0.0111} & \textbf{0.1864} \\
       \cmidrule{3-12}
       & ($\emph{\textbf{Synthetic}}$) \\
       & Full-CNN\cite{FullCNN_2023} & - & - & 0.5292 & - & - & - & - & - & 0.6751 & -\\
       & CRNN\cite{CRNN_2021} & 0.0568 & - & 0.3460 & - & 2.138 & 2.083 & - & - & - & - \\
       & CNN-MLP\cite{Srivastava2021BlindRP} & - & - & 0.1230 & - & - & - & - & - & 0.1198 & - \\
       & TAE-CNN\cite{Suradej} & 0.1293 & - & 0.9896 & 0.9643 & 5.559 & - & 25.09 & 0.0703 & - \\
       & RE-NET\cite{ZHENG2022110901} & - & - & 0.6624 & - & - & - & - & - & - & - \\
       & BERP (Ours) & \textbf{0.0041} & \textbf{0.1589} & \textbf{0.0219} & \textbf{0.0239} & \textbf{0.1991} & \textbf{0.2067} & \textbf{0.9783} & \textbf{0.0022} & \textbf{0.0300} & \textbf{0.4418} \\
       \hline
       & ($\emph{\textbf{Real}}$) \\
       \multirow{13}{*}{PCC $\uparrow$} 
       & Full-CNN\cite{FullCNN_2023} & - & - & 0.5248 & - & - & - & - & - & 0.1585 & - \\
       & CRNN\cite{CRNN_2021} & 0.9630 & - & 0.9423 & - & 0.9647 & 0.9579 & - & - & - & -  \\
       & CNN-MLP\cite{Srivastava2021BlindRP} & - & - & 0.9529 & - & - & - & - & - & 0.8566 & - \\
       & TAE-CNN\cite{Suradej} & - & - & - & - & - & - & - & - & - & - \\
       & RE-NET\cite{ZHENG2022110901} & - & - & 0.1189 & - & - & - & - & - & - & - \\
       & BERP (Ours) & \textbf{0.9960} & \textbf{0.9964} & \textbf{0.9973} & \textbf{0.9979} & \textbf{0.9964} & \textbf{0.9950} & \textbf{0.9947} & \textbf{0.9955} & \textbf{0.9819} & \textbf{0.9716} \\
       \cmidrule{3-12}
       & ($\emph{\textbf{Synthetic}}$) \\
       & Full-CNN\cite{FullCNN_2023} & - & - & 0.9400 & - & - & - & - & - & 0.5691 & - \\
       & CRNN\cite{CRNN_2021} & 0.9529 & - & 0.9434 & - & 0.9620 & 0.9511 & - & - & - & -\\
       & CNN-MLP\cite{Srivastava2021BlindRP} & - & - & 0.9870 & - & - & - & - & - & 0.9403 & - \\
       & TAE-CNN\cite{Suradej} & - & - & - & - & - & - & - & - & - & - \\
       & RE-NET\cite{ZHENG2022110901} & - & - & 0.7221 & - & - & - & - & - & -  & - \\
       & BERP (Ours) & \textbf{0.9956} & \textbf{0.9962} & \textbf{0.9987} & \textbf{0.9983} & \textbf{0.9946} & \textbf{0.9931} & \textbf{0.9919} & \textbf{0.9975} & \textbf{0.9737} & \textbf{0.9565} \\
    \specialrule{.2em}{.1em}{.1em} 
    \end{tabular}
    \label{tab:room_param_accu}
\end{table*}

\begin{table*}[!t]
    \centering
    \caption{Comparison between JND of RAPs\cite{ISO3382, ISRA, Alcons} and SD of estimation error on real and synthetic data.}
    \begin{tabular}{cccccccccc}
     \specialrule{.2em}{.1em}{.1em} 
       & & STI & $\%\text{AL}_{\text{cons}}$ & $T_{60}$ & EDT & $C_{80}$ & $C_{50}$ & $D_{50}$ & $T_s$ \\
        \specialrule{.1em}{.05em}{.05em}
        JND & & 0.03 & 3\% & 5\% & 5\% & 1 dB & 1 dB & 5\% & 10 ms \\
         \hline
         \multirow{2}{*}{SD}
         & Real & 0.01 & 4.9\% & 4\% & 4\% & 0.37 dB & 0.40 dB & 2\% & 5 ms \\
         & Synthetic & 0.01 & 6.7\% & 9\% & 8\% & 0.55 dB & 0.53 dB & 2\% & 7 ms \\
     \specialrule{.2em}{.1em}{.1em} 
    \end{tabular}
    \label{tab:jnd}
\end{table*}

\begin{table}[!t]
  \centering
  \caption{Evaluation results for occupancy level $N$ when four types of featurizers are used. \textbf{Bold} indicates the best result.}
  \begin{tabular}{ccccc}
     \specialrule{.2em}{.1em}{.1em} 
    & BERP-variants & macro-averaged MAE $\downarrow$ & MAE $\downarrow$ \\ 
    \specialrule{.1em}{.05em}{.05em}
     & MFCC & \textbf{1.107} & \textbf{0.3560} \\
     & Mel spectrogram & 1.252 & 0.3608 \\
     & Gammatonegram & 1.255 & 0.3735 \\
     & Spectrogram & 1.617 & 0.3941 \\
      \specialrule{.2em}{.1em}{.1em} 
  \end{tabular}
  \label{tab:occu_eval}
  \vspace{-1.5em}
\end{table}
\begin{table}[!t]
  \centering
  \caption{Evaluation results for each occupancy level $N$ by using the MAE and precision, along with count of time frames.}
  \begin{tabular}{ccccc}
     \specialrule{.2em}{.1em}{.1em} 
    & $N$ & MAE $\downarrow$ & F1 score [\%] $\uparrow$ & Count of time frames \\ 
    \specialrule{.1em}{.05em}{.05em}
     & 0 & 0.1049 & 87 & 1001 \\
     & 1 & 0.0825 & 92 & 9102 \\
     & 2 & 0.2440 & 76 & 6282 \\
     & 3 & 0.3897 & 61 & 4421 \\
     & 4 & 0.5314 & 51 & 3334 \\
     & 5 & 0.6692 & 42 & 2301 \\
     & 6 & 0.7000 & 41 & 1480 \\
     & 7 & 0.8291 & 35 & 925 \\
     & 8 & 0.9885 & 30 & 525 \\
     & 9 & 1.241 & 24 & 315 \\
     & 10 & 1.368 & 21 & 171 \\
     & 11 & 1.525 & 20 & 120 \\
     & 12 & 1.740 & 14 & 50 \\
     & 13 & 1.736 & 24 & 19 \\
     & 14 & 1.571 & 31 & 7 \\
     & 15 & 4.000 & 0 & 4 \\
      \specialrule{.2em}{.1em}{.1em} 
  \end{tabular}
  \label{tab:occu_eval_per_class}
  \vspace{-1.5em}
\end{table}

\begin{table*}[!t]
    \centering
    \caption{Results of an ablation study concerning separate estimation. We compared the evaluation results from a joint model and multiple separate models for each room parameter.}
    \begin{tabular}{ccccccccccccc}
     \specialrule{.2em}{.1em}{.1em} 
        & & STI & $\%\text{AL}_{\text{cons}}$ & $T_{60}$ & EDT & $C_{80}$ & $C_{50}$ & $D_{50}$ & $T_s$ & $V$ & $D$ \\
        & & & & [$s$] & [$s$] & [dB] & [dB] & [\%] & [$s$] & [$\log_{10}(m^3)$] & [$m$] \\
       \specialrule{.1em}{.05em}{.05em} 
       \multirow{2}{*}{MAE $\downarrow$} &
       Separate & 0.0094 & 0.2308 & 0.0394 & 0.0426 & 0.2858 & 0.2904 & 1.743 & 0.0054 & 0.0704 & 1.114 \\
       & Joint &  0.0043 & 0.1585 & 0.0206 & 0.0232 & 0.2066 & 0.2101 & 0.9915 & 0.0023 & 0.0261 & 0.3845 \\
       \hline
       \multirow{2}{*}{PCC $\uparrow$} 
       & Separate & 0.9897 & 0.9957 & 0.9921 & 0.9952 & 0.9924 & 0.9903 & 0.9872 & 0.9888 & 0.9741 & 0.8595 \\
       & Joint & 0.9958 & 0.9964 & 0.9984 & 0.9982 & 0.9949 & 0.9936 & 0.9925 & 0.9971 & 0.9756 & 0.9597 \\
       \specialrule{.2em}{.1em}{.1em} 
       \vspace{-1.5em}
    \end{tabular}
    \label{tab:sep_pipeline}
\end{table*}

\section{Experiments}
\label{sec:expres}
\subsection{Experimental Setup}
\label{ssec:expsetup}
\textbf{Training strategy.} For the dataset for the unified module, we use 47,430 synthetic and 11,160 real reverberant audio clips. We randomly allocate 43,430 synthetic clips and 10,044 real clips for training, and 2,000 synthetic clips and 558 real clips for validation. The trained models are tested on separate sets of 2,000 synthetic clips and 558 real clips. For the occupancy module, we randomly split the total of 47,430 audio clips into 43,430 training, 2000 validation, and 2000 test datasets. The padding mask is deployed to ensure that the model learns only the valid information across each minibatch. We used weighted random sampling on each batch to address the imbalance of synthetic and real samples in the training data. The RAdam optimizer with $\mathcal{L}2$ regularization is used \cite{RAdam}, which possesses a learning rate warmup functionality without the risk of underfitting the regression tasks. We utilize cosine-annealing and tri-stage learning rate schedulers for the unified and the occupancy modules, respectively, to facilitate the convergence of the models toward the global optimums. We set a batch size of $24$ on an H100 GPU. Given the wide range of room volumes spanning from $40$ to $9000$ $m^3$, we apply logarithmic scaling to compress them, stabilizing the training process and improving the robustness of the models. Before passing through the featurizer, the amplitudes of the input signals are normalized via peak normalization.
\begin{table*}[!t]
    \centering
    \caption{Results of an ablation study concerning auxiliary task learning. ``-xRPs" means the variant of estimating x room parameters. ``BERP" represents the standard BERP model. ``-" indicates no data available or no measurement performed.}
    \begin{tabular}{ccccccccccccc}
    \specialrule{.2em}{.1em}{.1em} 
        & & STI & $\%\text{AL}_{\text{cons}}$ & $T_{60}$ & EDT & $C_{80}$ & $C_{50}$ & $D_{50}$ & $T_s$ & $V$ & $D$ \\
        & Model & & & [$s$] & [$s$] & [dB] & [dB] & [\%] & [$s$] & [$\log_{10}(m^3)$] & [$m$] \\
        \specialrule{.1em}{.05em}{.05em}
        \multirow{10}{*}{MAE $\downarrow$} 
        & BERP-2RPs & - & - & 0.0336 & - & - & - & - & - & 0.0262 & - \\
        & BERP-4RPs & 0.0075 & - & 0.0301 & - & 0.2632 & 0.2621 & - & - & - & - &  \\
        & BERP-6RPs & 0.0077 & - & 0.0313 & 0.0351 & 0.2765 & - & 1.360 & 0.0033 & - & - \\
        & BERP & 0.0043 & 0.1585 & 0.0206 & 0.0232 & 0.2066 & 0.2101 & 0.9915 & 0.0023 & 0.0261 & 0.3845 \\
        & Full-CNN-10RPs & 1.266 & 3.179 & 1.390 & 1.487 & 3.596 & 2.771 & 129.9 & 1.970 & 0.8015 & 3.737 \\
        & CRNN-10RPs & 0.0662 & 2.267 & 0.3998 & 0.4398 & 2.152 & 2.080 & 12.03 & 0.0382 & 0.3168 & 2.629  \\
        & CNN-MLP-10RPs & 0.0186 & 0.8237 & 0.1361 & 0.1312 & 0.7932 & 0.7802 & 3.571 & 0.0109 & 0.1477 & 1.437 \\
        & TAE-CNN-8RPs & 0.1402 & 6.729 & 0.9515 & 0.9440 & 5.922 & 5.453 & 26.42 & 0.0694 & - & - \\
        & RE-NET-10RPs & 2.542 & 5.947 & 1.972 & 2.019 & 4.670 & 3.918 & 254.1 & 3.080 & 0.6910 & 4.419 \\
        \hline
        \multirow{10}{*}{PCC $\uparrow$} 
        & BERP-2RPs & - & - & 0.9979 & - & - & - & - & - & 0.9709 & -  \\
        & BERP-4RPs & 0.9919 & - & 0.9975 & - & 0.9938 & 0.9928 & - & - & - & - \\
        & BERP-6RPs & 0.9925 & - & 0.9976 & 0.9971 & 0.9935 & - & 0.9905 & 0.9957 & - & - \\
        & BERP & 0.9958 & 0.9964 & 0.9984 & 0.9982 & 0.9949 & 0.9936 & 0.9925 & 0.9971 & 0.9756 & 0.9597 \\
        & Full-CNN-10RPs & 0.1474 & 0.7991 & 0.6283 & 0.1747 & 0.7761 & 0.7575 & 0.6368 & 0.1683 & 0.1445 & 0.4386 \\
        & CRNN-10RPs & 0.9498 & 0.9561 & 0.9511 & 0.9552 & 0.9495 & 0.9372 & 0.9043 & 0.9267 & 0.6813 & 0.7250\\
        & CNN-MLP-10RPs & 0.9756 & 0.9732 & 0.9798 & 0.9767 & 0.9723 & 0.9694 & 0.9701 & 0.9745 & 0.9200 & 0.9076 \\
        & TAE-CNN-8RPs & - & - & - & - & - & - & - & - & - & - \\
        & RE-NET-10RPs & 0.0156 & 0.2000 & 0.2261 & 0.2245 & 0.1352 & 0.1621 & 0.1233 & 0.1596 & 0.0119 & 0.0103 \\
        \specialrule{.2em}{.1em}{.1em} 
        \vspace{-1.5em}
    \end{tabular}
    \label{tab:ablation_aux}
\end{table*}

\textbf{Featurizer configuration.} We set a uniform configuration for all spectrogram-variant featurizers. They each contain the same 128 frequency bins, Hanning windowing with a size of $1024$, and a $75\%$ overlapping rate.

\textbf{Baselines.} In our comparative experiments, we evaluated the performance of our proposed method with five baseline architectures that are renowned in the domain of room parameter estimation amidst background noise: Full-CNN \cite{FullCNN_2023, FullCNN_2019, FullCNN_2018}, CRNN \cite{CRNN_2021, CRNN_2020}, CNN-MLP\cite{Srivastava2021BlindRP}, TAE-CNN \cite{Suradej}, and RE-NET \cite{ZHENG2022110901}. 
We adopt the configurations of the Full-CNN and CRNN architectures based on their latest variants, as described in \cite{FullCNN_2023} and \cite{CRNN_2021}, respectively. We retrained and compared all baselines on the room parameters they are designed to estimate to ensure fair comparisons. All the baselines were sufficiently trained for convergence.

\subsection{Evaluation Metrics}
\label{ssec:evalmetrics}
We use the mean absolute error (MAE) and the Pearson correlation coefficient (PCC) as evaluation metrics for the blind estimation of RAPs and RGPs. The MAE provides a direct measure of the scale of the average estimation error, and the PCC is introduced to quantify the agreement of the estimated and ground-truth values. When estimating occupancy levels, we use the macro-averaged MAE and MAE to quantify the Euclidean distance between the estimated and ground-truth occupancy sequences for each occupancy level, given the imbalanced class distribution of the data.

\subsection{Evaluation Results}
\label{ssec:expresults}
The evaluation results are presented in three subsections. Section\,\ref{sssec:eval_rap_rgp} compares BERP with the baselines for RAPs and RGPs, Section\,\ref{sssec:eval_occu_lv} presents the occupancy level results, and Section\,\ref{sssec:ablation} details the ablation studies.
\subsubsection{Evaluation of RAPs and RGPs}
\label{sssec:eval_rap_rgp}
First, we evaluated four BERP variants with different featurizers to determine which type is most effective and to clarify the contribution of auditory-inspired time-frequency representations. The comparison results on the real dataset are shown in Table\,\ref{tab:feat_compare}. Among four featurizers, MFCC performed the best, highlighting the potential benefits of auditory-inspired time-frequency representations. The mel spectrogram and Gammatonegram proved less effective for this task. We chose the best-performing MFCC variant for comparison with the baselines. Table\,\ref{tab:room_param_accu} presents the estimation results of BERP and the baseline methods on both real and synthetic datasets. All the models were evaluated via the same dataset described in Section\,\ref{sssec:noisyReverbData} with the data segmentation settings in Section\,\ref{ssec:expsetup}. To assess the adaptability of the models to both simulated and real acoustical environments, evaluations were carried out separately on real and synthetic datasets. Notably, the baseline methods rarely include evaluations of real recordings, making this comparison particularly significant. The results show that BERP significantly outperforms SOTA architectures in terms of the MAE in both simulated and real settings. The PCC also revealed that the estimated room parameters strongly agreed with the ground truths, indicating the effectiveness of our model. The results of the MAE and PCC consistently indicate the superiority of the proposed architecture for jointly estimating RAPs and RGPs, achieving significantly better performance than the baselines in terms of the room parameters they are designed to estimate. Additionally, our approach can blindly estimate ten room parameters simultaneously with satisfactory accuracy, a capability that the baselines fail to achieve.

The estimation accuracy of RAPs in relation to subjective perception can be quantified by the just noticeable difference (JND), which is the subjective limen to changes in a given RAP. To evaluate this accuracy, we compare the standard derivation (SD) of the estimation error with the JND of each RAP. The JND and SD of all RAPs are shown in Table\,\ref{tab:jnd}. We chose the best-performing MFCC featurizer. The results indicate that our approach achieves satisfactory accuracy in representing the subjective perception of each RAP.

\subsubsection{Evaluation of instantaneous occupancy level}
\label{sssec:eval_occu_lv}
Table\,\ref{tab:occu_eval} presents the evaluation results of the occupancy level estimation. As noted in Section\,\ref{sec:related_work}, to our knowledge, no recorded speech dataset is available for blind estimation of occupancy levels. Hence, we rely solely on synthetic data for this study. The creation of a real dataset is left for future work. Among the four featurizers, the MFCC performed best, underscoring the benefit of an auditory-inspired time-frequency representation. We then compared the MAEs and F1 scores for each occupancy level in Table\,\ref{tab:occu_eval_per_class}, using the top-performing MFCC. The results show that the F1 score used in previous studies \cite{occu_hvac, Chen2017AnIO} is insufficient, as it treats any inexact prediction as entirely incorrect and fails to reflect how closely the estimated count aligns with the ground truth.
\subsubsection{Ablation Study}
\label{sssec:ablation}
\textbf{Separate estimation.}
This ablation study aims to verify our intuition that a joint estimation leads to better performance than separate estimations. Specifically, we trained separate models for each RAP and RGP and compared their performance with that of the joint estimation model. In total, we trained ten individual models, one for each RAP and RGP. The MFCC featurizer is used to train all the models. Table\,\ref{tab:sep_pipeline} compares the performance of separate and joint models. The results confirm that the joint estimation model consistently outperforms the separate models for each room parameter, substantiating that mutual relatedness among RAPs and RGPs may benefit knowledge transfer and lead to more sufficient learning. Moreover, estimations of the STI, $D_{50}$, $T_s$, $V$, and $D$ are harder tasks that benefit significantly from the joint estimation strategy. 

\textbf{Auxiliary task learning.} We aim to verify the hypothesis that adding more auxiliary tasks can improve overall performance, given the mutual relatedness among RAPs and RGPs (see Sections\,\ref{sec:intro} and \ref{sec:related_work}). In this case, rather than estimating two, four, or six room parameters as baselines do, introducing additional room parameters (adding auxiliary tasks) may lead to better performance of one or more primary tasks. In this ablation study, we trained BERP variants designed to estimate the same room parameters that the baselines estimate and compare them with the standard BERP model. The MFCC featurizer is used here to train all the models. Table\,\ref{tab:ablation_aux} shows the evaluation results, which confirm that incorporating more tasks leads to better performance.

In addition, we intend to highlight both the advantages of our architecture and the limitations of existing methods when jointly estimating numerous room parameters. In this case, we trained the baseline models to predict all RAPs and RGPs. The results in Table\,\ref{tab:ablation_aux} also indicate that the  performance of the baselines significantly deteriorates on room parameters they were designed to estimate when many parameters are jointly estimated, seeming to support that their model design limits their ability to handle multiple tasks effectively. 
\begin{table*}[t!]
    \centering
    \caption{Results of an ablation study concerning hybrid vs. end-to-end approaches.}
    \begin{tabular}{ccccccccccccc}
    \specialrule{.2em}{.1em}{.1em} 
         & & STI & $\%\text{AL}_{\text{cons}}$ & $T_{60}$ & EDT & $C_{80}$ & $C_{50}$ & $D_{50}$ & $T_s$ & $V$ & $D$ \\
        & Architecture & & & [$s$] & [$s$] & [dB] & [dB] & [\%] & [$s$] & [$\log_{10}(m^3)$] & [$m$] \\
        \specialrule{.1em}{.05em}{.05em}
        \multirow{2}{*}{MAE $\downarrow$} 
         & Hybrid & 0.0414 & 3.031 & 0.0489 & 0.2857 & 2.640 & 3.031 & 13.10 & 0.0490 & 0.0304 & 0.04509 \\
         & Proposed & 0.0045 & 0.1647 & 0.0199 & 0.0225 & 0.2130 & 0.2164 & 1.001 & 0.0023 & 0.0271 & 0.4822 \\
         \hline
         \multirow{2}{*}{PCC $\uparrow$}
         & Hybrid & 0.9532 & 0.9149 & 0.9973 & 0.9873 & 0.9034 & 0.8411 & 0.8376 & 0.9752 & 0.9688 & 0.9397 \\
         & Proposed & 0.9956 & 0.9963 & 0.9989 & 0.9987 & 0.9940 & 0.9924 & 0.9912 & 0.9976 & 0.9748 & 0.9606 \\
         \specialrule{.2em}{.1em}{.1em}
    \end{tabular}
    \label{tab:ablation_hybrid_vs_e2e}
\end{table*}
\vspace{-0.2em}
\begin{table*}[t!]
    \centering
    \caption{Results of an ablation study concerning simulated vs real RIR.}
    \begin{tabular}{ccccccccccccc}
    \specialrule{.2em}{.1em}{.1em} 
         & & STI & $\%\text{AL}_{\text{cons}}$ & $T_{60}$ & EDT & $C_{80}$ & $C_{50}$ & $D_{50}$ & $T_s$ & $V$ & $D$ \\
        & RIR & & & [$s$] & [$s$] & [dB] & [dB] & [\%] & [$s$] & [$\log_{10}(m^3)$] & [$m$] \\
        \specialrule{.1em}{.05em}{.05em}
        \multirow{2}{*}{MAE $\downarrow$} 
         & Simulated & 0.0558 & 2.463 & 0.5850 & 0.6703 & 4.798 & 4.788 & 19.72 & 0.0680 & 0.9197 & 4.568 \\
         & Real & 0.0045 & 0.1647 & 0.0199 & 0.0225 & 0.2130 & 0.2164 & 1.001 & 0.0023 & 0.0271 & 0.4822 \\
         \hline
         \multirow{2}{*}{PCC $\uparrow$}
         & Simulated & 0.8620 & 0.8620 & 0.8988 & 0.9162 & 0.8720 & 0.8507 & 0.8535 & 0.9090 & 0.3162 & 0.3354 \\
         & Real & 0.9956 & 0.9963 & 0.9989 & 0.9987 & 0.9940 & 0.9924 & 0.9912 & 0.9976 & 0.9748 & 0.9606 \\
         \specialrule{.2em}{.1em}{.1em}
    \end{tabular}
    \label{tab:ablation_sim_vs_real}
\end{table*}
\begin{table*}[t!]
    \centering
    \caption{Results of an ablation study concerning without the parametric predictor. We simply replace the PP with an FC layer for each room parameter.}
    \begin{tabular}{ccccccccccccc}
    \specialrule{.2em}{.1em}{.1em} 
    & & STI & $\%\text{AL}_{\text{cons}}$ & $T_{60}$ & EDT & $C_{80}$ & $C_{50}$ & $D_{50}$ & $T_s$ & $V$ & $D$ \\
        & BERP-variants & & & [$s$] & [$s$] & [dB] & [dB] & [\%] & [$s$] & [$\log_{10}(m^3)$] & [$m$] \\
        \specialrule{.1em}{.05em}{.05em}
        \multirow{2}{*}{MAE $\downarrow$} 
        & BERP w/o PP & 0.0060 & 0.1714 & 0.0269 & 0.0290 & 0.2325 & 0.2344 & 1.087 & 0.0043 & 0.0381 & 0.3677 \\
        & BERP & 0.0043 & 0.1585 & 0.0206 & 0.0232 & 0.2066 & 0.2101 & 0.9915 & 0.0023 & 0.0261 & 0.3845 \\
        \hline
        \multirow{2}{*}{PCC $\uparrow$} 
        & BERP w/o PP & 0.9956 & 0.9963 & 0.9985 & 0.9981 & 0.9954 & 0.9943 & 0.9939 & 0.9950 & 0.9825 & 0.9697 \\
        & BERP & 0.9958 & 0.9964 & 0.9984 & 0.9982 & 0.9949 & 0.9936 & 0.9925 & 0.9971 & 0.9756 & 0.9597 \\
        \specialrule{.2em}{.1em}{.1em} 
    \end{tabular}
    \vspace{-1.2em}
    \label{tab:ablation_wopp}
\end{table*}

\textbf{Hybrid vs End-to-end.} We also explored a hybrid approach inspired by \cite{blindEst_wang, Suradej}. Rather than predicting eight RAPs directly, we use a fewer-parameter stochastic RIR modeling and predict its parameters, which are then used to reconstruct the RIR and derive RAPs. In contrast, we estimate RGPs end-to-end. This method aims to simplify the model by reducing the number of parameters to be estimated, while improving the consistency among RAPs by deriving from the same RIR. Owing to page limitation, more details for the hybrid approach are provided in Sections II (architecture) and III (experimental results) of the supplementary material (see attached material below). Table\,\ref{tab:ablation_hybrid_vs_e2e} shows the comparisons between the hybrid and end-to-end architectures, which indicate that the end-to-end architecture significantly outperforms the hybrid architecture. We suspect that the limitations of the hybrid approach stem from the mismatch of the RIR model with real RIRs, which likely prevents it from capturing all the complexities of real RIRs, thereby reducing the effectiveness of the hybrid approach. In addition, this hybrid approach might benefit less from auxiliary task learning, considering that the parameters of the RIR model and $V$ and $D$ are mutually independent.

\textbf{Simulated vs Real RIR.} This ablation study assesses the effectiveness of using simulated versus real RIRs for synthesizing reverberant signals, as discussed in Section\,\ref{sec:related_work}. We used an open-sourced \emph{pyroomacoustic} RIR simulation algorithm \cite{pyroomacoustic} to generate simulated RIRs on the basis of the same meta-information (room volume, $T_{60}$, room dimensions, source and microphone positions, etc.) of real RIRs used to train the standard BERP model. Next, we applied the data synthesis pipeline described in Section\,\ref{sssec:noisyReverbData} to synthesize the observed reverberant signals via these simulated RIRs. We then trained two models: one using synthetic data derived from simulated RIRs and another using data derived from real RIRs. The evaluation results shown in Table\,\ref{tab:ablation_sim_vs_real} indicate that using real RIRs to compile synthetic data offers superior performance. Notably, when generating these simulated RIRs, we did not completely replicate the detailed room geometries, material configurations, or configurations of interior objects (e.g., furniture and decorations) of real rooms where the real RIRs were measured, since accessing this meta-information is nearly impossible. To fully replicate the exact environments where the real RIRs were measured, one would have to reproduce precise room geometries and layouts and object arrangements. However, this is nearly an impractical task, given that obtaining 3D scans of various room architectures (e.g., hotels, churches, and concert halls) is often unavailable and recreating interior object settings is impractical, particularly at a large scale. Consequently, we offer a simplified baseline that still provides a meaningful insight into the realism of room simulation for training purposes.

\textbf{Without the parametric predictor.} To dissect the contributions of the RFE and PP to the overall performance of BERP, we conducted an ablation study. We consistently used the MFCC featurizer. Table\,\ref{tab:ablation_wopp} compares the results obtained using solely the RFE with those achieved by using the full architecture equipped with the PP. The results show that the RFE contributes significantly to the performance of BERP.

\section{Conclusion}
\label{sec:conclu}
\textbf{Discussion.} This subsection recaps key points related to how BERP's model design leads to its superior performance. 
On the one hand, combining self-attention with convolutions enables BERP to capture acoustic features globally and locally, leading to more robust and task-generalizable shared representations over different tasks. Furthermore, by sharing representations across all tasks, the model benefits from inductive bias learning \cite{Baxter2011} and the regularization effect\cite{Wang2022, ruder2017overviewmultitasklearningdeep, liu-etal-2019-multi}, leading to faster convergence, especially for harder tasks or those with fewer training examples \cite{mccann2018naturallanguagedecathlonmultitask, liu-etal-2019-multi}. On the other hand, giving each task its own parameters mitigates gradient conflicts at the output layer, which is an issue that often arises in the single architecture baselines adopted. BERP can also predict the instantaneous occupancy level $N$ in a short time window that baselines do not offer. BERP further distinguishes itself as a sequence-to-sequence model, allowing it to handle arbitrary length inputs without explicit length alignment, an advantage over most existing approaches. 

\textbf{Conclusion.} We propose BERP, a  blind estimation approach that compromises two separately trained modules: one that jointly estimates RAPs and RGPs, and another that predicts the occupancy level, each using its own single-channel speech input. For RAP and RGP estimation, we employ a shared encoder paired with multiple task-specific predictors, whereas we use an identical encoder coupled with a classification head to estimate occupancy levels over time. They share a similar design philosophy and offer a more complete solution for the blind estimation of room acoustics. Our approach fills the current gap by providing a universal solution that addresses multiple room parameters and occupancy levels. Evaluations on both real and synthetic data demonstrate that BERP significantly outperforms existing methods, achieving SOTA results by estimating eleven room parameters simultaneously for the first time. These findings underscore BERP’s potential across a wide range of applications in room acoustics, hearing aids, communication, and human–machine interactions. 

\textbf{Limitations and Future work.} Except for occupancy level estimation, BERP assumes a dynamically moving, single-source speech signal as the observed input. Future research will aim to address the blind estimation of RAPs and RPPs for multisource speech signals. Furthermore, this work assumes omnidirectional sources and receivers, which may limit its application, since \cite{Srivastava2022} showed that directivity has a great impact on performance. We plan to explore this issue in future work. Another noted gap stems from the data generation for occupancy level estimation. We make certain simplifications and limitations that may not fully reflect real-world settings; addressing these issues, along with the collection of datasets of real recordings, is left for future work.

\vspace{-0.5em}
\section*{Acknowledgments}
We appreciate the great help from Jianan Chen for this work.

\begingroup
\let\itshape\upshape

\bibliographystyle{ieeetr}
\bibliography{manuscript}
\endgroup

\end{document}


\setlength{\skip\footins}{6pt}
\title{Supplementary Material for ``BERP: A Blind Estimator of Room Parameters for Single-Channel Noisy Speech Signals"}

\markboth{Journal of \LaTeX\ Class Files,~Vol.~xx, No.~xx, xx ~xxxx}%
{How to Use the IEEEtran \LaTeX \ Templates}

\maketitle

\begingroup\renewcommand\thefootnote{\textsection}
\endgroup
\section{Introduction}
\begin{figure*}[b!]
    \centering
    \includegraphics[width=0.88\textwidth]{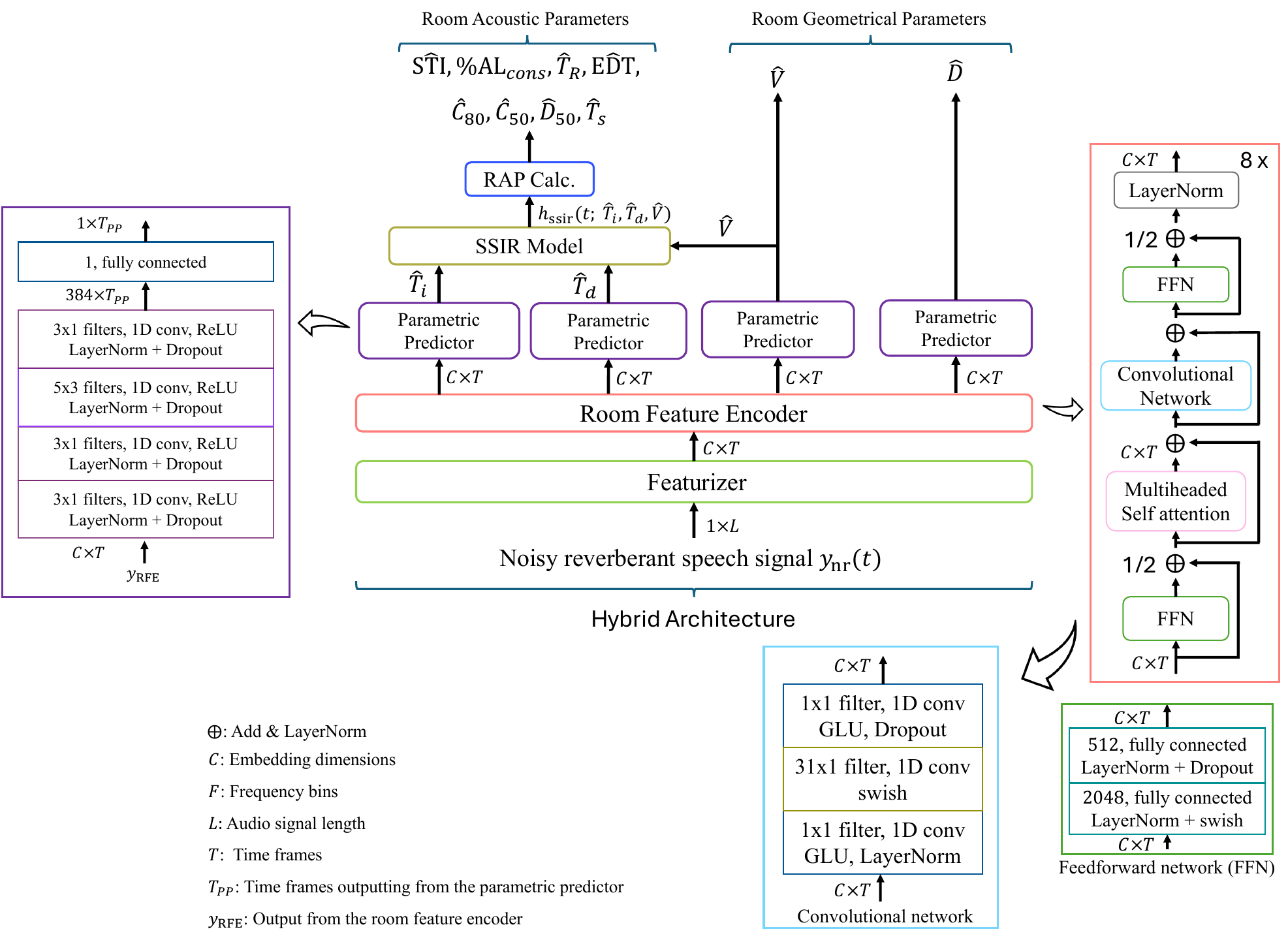}
    \caption{Overview of the hybrid architecture for jointly estimating RAPs and RGPs. The input is the observed noisy reverberant speech signal $y_{nr}(t)\in\mathbb{R}^{1\times L}$. Details of the SSIR model are provided in Section\,\ref{ssec:ssir}, and the architecture of the room feature encoder and parametric predictor is described in Section V-B of the main manuscript. During inference, the estimated parameters $\hat{T_i}$, $\hat{T_d}$, $\hat{V}$, and $\hat{D}$ are summed and averaged over the time frames $T_{PP}$ to yield single scalar values. The values $\hat{T_i}$, $\hat{T_d}$, and $\hat{V}$ are then input to the SSIR model to simultaneously derive the RAPs.} 
    \label{fig:block_diagram}
\end{figure*}

Figure\,S\ref{fig:block_diagram} illustrates the hybrid approach for the joint estimation of RAPs and RGPs. Unlike the proposed BERP framework that estimates all parameters end-to-end, this hybrid design uses the neural network to predict the parameters of a stochastic RIR energetic model (``SSIR Model" in the Fig.\,S\ref{fig:block_diagram}) with far fewer parameters and then derive RAPs from the generated RIR, while RGPs are directly estimated from the neural networks. This hybrid way aims to simplify the model by reducing the number of parameters to be estimated while improving consistency among RAPs, since the number of parameters of the stochastic RIR model is much fewer than that of RAPs and all RAPs can be derived from the same RIR. Given that the neural network components in this hybrid approach are identical to those in the end-to-end model, we refer readers to the main manuscript for further details.

The remainder of this supplementary material is organized as follows. Section\,\ref{sec:hybrid} details the key differences between this hybrid approach and the proposed method in the main manuscript. Section\,\ref{sec:expres} presents the experimental results of the hybrid approach on real and synthetic data.

\section{Hybrid Approach}
\label{sec:hybrid}

\subsection{Sparse Stochastic Impulse Response Model}
\label{ssec:ssir}
An RIR can be modeled stochastically using only a few control parameters, making this approach both simple and effective for blind room parameter estimation\cite{blindEst_wang, Suradej, Unoki}. For example, Schroeder’s RIR model approximates an unknown RIR as an ideal exponential decay\cite{Schroeder1962, schroeder1981}:
\begin{equation}
    h_{\rm{Sch}}(t) = e_{h}(t)\textbf{c}(t) = a \exp\bigg(-\frac{6.9}{T_{60}}\bigg)\textbf{c}_{\rm{Sch}}(t),
\end{equation}
where $e_h(t)$ is the temporal envelope, $\textbf{c}_{\rm{Sch}}(t)\sim \mathscr{N}(0,1)$ is a white Gaussian noise carrier, and $a$ is a gain factor. Although this ideal exponential decay function is adequate for geometrically simple rooms, it does not accurately represent more complex environments characterized by irregular geometries and furnishings, because it fails to model the onset transition of the measured RIR \cite{Unoki2017APSIPA, Suradej, blindEst_wang}.

To address this limitation, an extended RIR model was proposed \cite{Unoki2017APSIPA, Suradej, blindEst_wang}:
\begin{equation}
    h_{\rm{ext}}(t)=e_h(t)\textbf{c}(t)=
    \begin{cases}
        a e^{-6.9/T_h}\textbf{c}_{\rm{ext}}(t),\hspace{15pt} t \in [0, T_h) \\
        a e^{6.9/T_t}\textbf{c}_{\rm{ext}}(t),\hspace{10pt} t\geq [T_h, T]
    \end{cases}
    \label{eq:ext_RIR}
\end{equation}
where $T_h$ and $T_t$ control the rising and decaying envelopes, $T$ is the total RIR duration, and $\textbf{c}_{\rm{ext}}(t)\sim \mathscr{N}(0,1)$ is a centered Gaussian process. $T$ is the time length of the RIR. Although this extended model better approximates real RIRs by modeling the onset, using a centered Gaussian process for the fine structure of the onset contradicts the sparse statistical nature observed in real RIR onsets \cite{Traer2016}.

Further insights come from Badeau’s unified stochastic framework based on the source image principle \cite{Badeau2018, Badeau_ssir, kuttruff2016room}, which shows that image sources (reflections) follow a uniform Poisson distribution that remains invariant with respect to the positions of the source and receiver. Polack \cite{POLACK1993_ssir} similarly demonstrated that this Poisson distribution is independent of room geometry. Moreover, Traer and McDermott \cite{Traer2016_ssir} found that while the late reverberation of an RIR is Gaussian, the onset is non-Gaussian. This discrepancy between conventional stochastic models and real RIR statistics can lead to inaccurate RAP estimation.
\begin{table*}[!t]
    \centering
    \caption{Accuracy (MAE) of derived RAPs from the extended and SSIR RIR model with respect to real RIRs.}
    \begin{tabular}{ccccccccccc}
     \specialrule{.2em}{.1em}{.1em} 
        & & STI & $\%\text{AL}_{\text{cons}}$ & $T_{60}$ & EDT & $C_{80}$ & $C_{50}$ & $D_{50}$ & $T_s$ \\
        & RIR model & & & [$s$] & [$s$] & [dB] & [dB] & [\%] & [$s$] \\
       \specialrule{.1em}{.05em}{.05em} 
       \multirow{2}{*}{MAE $\downarrow$} 
       & Extended & 0.0686 & 1.505 & 0.0735 & 0.1678 & 1.662 & 1.587 & 6.747 & 0.0207 \\
       & SSIR &  0.0685 & 1.501 & 0.0735 & 0.1678 & 1.661 & 1.586 & 6.742 & 0.0207 \\
       \specialrule{.2em}{.1em}{.1em} 
    \end{tabular}
    \label{tab:ablation_rir}
\end{table*}

Inspired by these findings, we propose a new stochastic energetic RIR model, called the sparse stochastic impulse response (SSIR) model, especially for a better approximation of the onset of RIRs. In the SSIR model, instead of using a Gaussian distribution, the onset transition is modeled as a uniform Poisson process with sparsity proportional to the room volume, which captures the sparse arrival times characteristic of early reflections in rooms. While the exponential decay is modeled as a centered Gaussian process, which captures the dense and complex arrival characteristic of late reverberation. Formally, the SSIR model is defined as:
\begin{equation}
\label{eq:SSIR_eq}
    h_{\rm{ssir}}(t) = 
    \begin{cases}
        h_{\rm{i}}(t) = a e^{6.9 t / T_i} c_i(t), \hspace{15pt} t \in [0, T_i) \\
        h_{\rm{d}}(t) = a e^{-6.9 t / T_d} c_d(t), \hspace{6pt} t \in [T_i, T]
    \end{cases}
\end{equation}
\begin{equation}
\label{eq:poisson}
    c_i(t) \sim {\mathscr{P}}(\lambda \lvert V \rvert) = \frac{\lambda^{\lfloor V \rfloor} \cdot e^{-\lambda}}{\lfloor V \rfloor! },
\end{equation}
\begin{equation}
\label{eq:gaussian}
    c_d(t) \sim \mathscr{N}(0, 1) = \frac{e^{-t^2/2}}{\sqrt{2\pi}}.
\end{equation}
\begin{figure}
    \centering
    \includegraphics[width=0.4\textwidth]{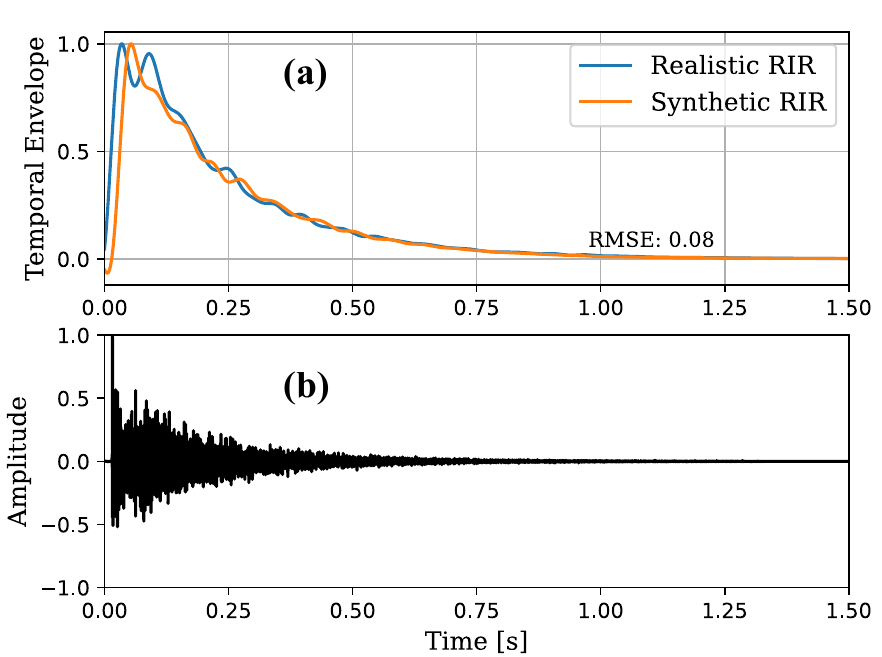}
    \caption{Fitting of the SSIR model to the realistic RIR. (a) The temporal envelope of the realistic and synthetic RIRs. (b) The corresponding realistic RIR.}
    \label{fig:SSIR}
\end{figure}
Here, $T_i$ and $T_d$ control the rising and falling temporal envelopes, respectively. Here, $c_i(t)$ models the fine structure of the RIR onset. It is represented as a realization of a Poisson process, which captures the sparse arrival times characteristic of early reflections in rooms. The parameter $\lambda$ is set equal to $\mu$ (the expectation of the Poisson distribution), which is the mean of $T_i$ in our dataset of 0.0399. Similarly, $c_d(t)$ models the fine structure of the RIR decay. It is represented as a realization of a Gaussian white noise process with zero mean and unit variance ($\mathscr{N}(0,1)$). Figure\,S\ref{fig:SSIR} demonstrates the fit of the SSIR model to a real RIR. 

Table\,S\ref{tab:ablation_rir} presents the results of an ablation study comparing the extended RIR model and the proposed SSIR model in terms of their ability to accurately represent real RIRs for the purpose of RAP derivation. In this study, parameters for both models were fitted to real RIRs, and eight standard RAPs were computed from the resulting synthetic RIRs. We modeled real RIRs using both approaches and computed eight RAPs from each modeled RIR as well as from the actual RIRs. The fitness of each model was then assessed by comparing the mean absolute error between the ground-truth RAPs and those calculated from the modeled RIRs. The results indicate that both models achieve a comparable level of accuracy in this RAP derivation task, exhibiting only minor differences in MAE across the various parameters. While this particular evaluation focuses on RAP accuracy after fitting and shows similar performance, the SSIR model's distinct onset structure, based on Poisson process statistics, is designed to better capture the sparse nature of early reflections \cite{Traer2016, Badeau_ssir, Badeau2018}. This structural difference is theoretically motivated and forms a key component of the hybrid approach. 
\begin{table*}[t!]
    \centering
    \caption{Evaluation results of the hybrid approach on \emph{real} and \emph{synthetic} data.}
    \begin{tabular}{ccccccccccccc}
    \specialrule{.2em}{.1em}{.1em} 
         & & STI & $\%\text{AL}_{\text{cons}}$ & $T_{60}$ & EDT & $C_{80}$ & $C_{50}$ & $D_{50}$ & $T_s$ & $V$ & $D$ \\
        & & & & [$s$] & [$s$] & [dB] & [dB] & [\%] & [$s$] & [$\log_{10}(m^3)$] & [$m$] \\
        \specialrule{.1em}{.05em}{.05em}
        \multirow{3}{*}{MAE $\downarrow$} 
         & \emph{Real} & 0.0260 & 1.020 & 0.0379 & 0.1244 & 1.675 & 1.913 & 7.279 & 0.0283 & 0.0195 & 0.2957 \\
         \cmidrule{3-12}
        & \emph{Synthetic} & 0.0460 & 3.585 & 0.0509 & 0.3286 & 2.903 & 3.336 & 14.75 & 0.0540 & 0.0335 & 0.4917 \\
         \hline
         \multirow{3}{*}{PCC $\uparrow$}
         & \emph{Real} & 0.9733 & 0.9725 & 0.9953 & 0.9856 & 0.9214 & 0.8797 & 0.9121 & 0.9702 & 0.9688 & 0.9300 \\
         \cmidrule{3-12}
        & \emph{Synthetic} & 0.9491 & 0.9032 & 0.9980 & 0.9879 & 0.8995 & 0.8325 & 0.8189 & 0.9768 & 0.9679 & 0.9430 \\
         \specialrule{.2em}{.1em}{.1em}
    \end{tabular}
    \label{tab:hybrid_eval_results}
\end{table*}

\subsection{Problem Formulation}
\label{ssec:prob_form}
We optimize the following objective function:
\begin{multline}
     \underset{\boldsymbol{\theta}^{sh}, \boldsymbol{\theta}^{T_i}, \boldsymbol{\theta}^{T_d}, \boldsymbol{\theta}^{V}, \boldsymbol{\theta}^{D}}{\min} \mathcal{L}^{T_i}(f^{T_i}(y_{nr}(t);\boldsymbol{\theta}^{sh},\boldsymbol{\theta}_{T_i}^{t}) \\ 
    + \mathcal{L}^{T_d}(f^{T_d}(y_{nr}(t);\boldsymbol{\theta}^{sh},\boldsymbol{\theta}_{T_d}^{t}) \\ + \mathcal{L}^{V}(f^{V}(y_{nr}(t);\boldsymbol{\theta}^{sh},\boldsymbol{\theta}_{V}^{t}) + \mathcal{L}^{D}(f^{D}(y_{nr}(t);\boldsymbol{\theta}^{sh},\boldsymbol{\theta}_{D}^{t}).
    \label{eq:obj}
\end{multline}
$f: \mathbb{R}^{1 \times L} \rightarrow \mathbb{R}^{1}$ is the trainable mapping function between the observed signal $y_{nr}(t) \in \mathbb{R}^{1 \times L}$ and target parameters, including two parameters of SSIR model $T_i \in \mathbb{R}^{1}$ and $T_d \in \mathbb{R}^{1}$, room volume $V \in \mathbb{R}^{1}$ and sound source distance $D \in \mathbb{R}^{1}$. $\boldsymbol{\theta}^{sh}$ is the shared parameters of the neural network, rather $\boldsymbol{\theta}^{t}$ represents the respective task-specific parameters for $T_i$,$\cdots$,$D$. The room feature encoder (RFE) is used to represent $\boldsymbol{\theta}^{sh}$, while the parametric predictor (PP) represents $\boldsymbol{\theta}^{t}$ for each estimated parameter. This objective function optimizes $\boldsymbol{\theta}^{sh}$, $\boldsymbol{\theta}_{T_i}^{t}$, $\cdots$, $\boldsymbol{\theta}_{D}^{t}$ to obtain the estimated parameters $\hat{T_i}$, $\hat{T_d}$, $\hat{V}$, and $\hat{D}$. We refer the reader to Section\,V-B of the main manuscript for more details on the RFE and PP.

Then, eight RAPs can be derived from the generated RIRs using SSIR model as:
\begin{equation}
    \hat{\gamma} = f_{\text{calc}}(h_{\text{ssir}}(t; \hat{T_i}, \hat{T_d}, \hat{V}))
\end{equation}
where $h_{\text{ssir}}(t; \hat{T_i}, \hat{T_d}, \hat{V})$ is the generated RIR from the SSIR model using the estimated three parameters $\hat{T_i}$, $\hat{T_d}$, and $\hat{V}$. $\hat{\gamma}$ is the estimated RAP. $f_{\text{calc}}$ is the corresponding calculation methods of RAPs by using the RIR, defined in IEC 60286-16:2020 \cite{IEC60268} and ISO 3382-1:2009 \cite{ISO3382}. We refer readers to Section III of the main manuscript for more details on RAP calculation methods.

\subsection{Loss Function}
\label{ssec:lossfunc}
We use Huber loss \cite{Huber_loss} as $\mathcal{L}$ in Eq.\, (\ref{eq:obj}) for $T_i$,$\cdots$,$D$.
\begin{equation}
\label{eq:predictor_loss}
    \mathcal{L}(x, \hat{x}) =
    \begin{cases}
        \frac{1}{2} \sum_{t=1}^{T_{PP}} (x - \hat{x}_{t})^2, \hspace{19pt} \lvert x - \hat{x}_{t} \rvert \leq \delta \\
        \delta \sum_{t=1}^{T_{PP}} \lvert x - \hat{x}_{t} \rvert - \frac{1}{2}\delta^2, \hspace{4pt} \lvert x - \hat{x}_{t} \rvert > \delta
    \end{cases}
\end{equation}
where $x \in \mathbb{R}^{1}$ is the ground-truth parameter (i.e., $T_i$, $T_d$, $V$, and $D$) and $\hat{x}_t \in \mathbb{R}^{1}$ is the corresponding estimated one at each time frame $t$. $\delta$ is set to 1. $T_{PP}$ represents the total time frame output from the PP, as shown in Fig.\,S\ref{fig:block_diagram}.

\section{Experiments}
\label{sec:expres}
\subsection{Experimental Setup and Evaluation Metrics}
\label{ssec:expsetup}
\textbf{Training strategy.} Because the experimental setup is identical to that of the proposed approach in the main manuscript, we refer readers to Section\,VI-A of the main manuscript for additional details. Likewise, since we use the same evaluation metrics (mean absolute error (MAE) and Pearson correlation coefficient (PCC)), we do not repeat those details here.

\subsection{Results}
\label{ssec:expresults}
We evaluated the hybrid approach on real and synthetic data using the same dataset described in Section IV-C1 of the main manuscript with the data segmentation setting described in Section VI-A of the main manuscript. We used MFCC featurizer here. Table\,S\ref{tab:hybrid_eval_results} shows the estimation results of the hybrid approach for RAPs and RGPs, which indicates that the hybrid approach still achieves a reasonably good performance even though it underperforms the proposed end-to-end approach.

\bibliographystyle{ieeetr}
\bibliography{supple}